\documentclass[acmsmall,9pt]{acmart}

\usepackage{amsmath}
\usepackage{multirow}
\usepackage{pifont}
\usepackage{subfig}
\usepackage{graphicx}
\usepackage{colortbl}
\usepackage{balance}
\usepackage{xcolor}
\usepackage[space]{grffile}
\usepackage{lscape}
\usepackage{caption}
\usepackage{tablefootnote}

\AtBeginDocument{
  \providecommand\BibTeX{{
    \normalfont B\kern-0.5em{\scshape i\kern-0.25em b}\kern-0.8em\TeX}}}

\newcommand{\ourmethod}{\textit{AQuaMoHo}}
\newcommand{\CityA}{Durgapur}
\newcommand{\CityB}{Delhi}

\newcommand{\blue}{black}

\begin{document}

\title{\ourmethod: Localized Low-Cost Outdoor Air Quality Sensing over a Thermo-Hygrometer}

\author{Prithviraj Pramanik}%
\authornote{Both the authors have equal contributions.}
\authornote{Corresponding Author}
\email{prithvirajpramanik@yahoo.co.in}
\affiliation{
\department{Department of Computer Science \& Engineering}
\institution{NIT Durgapur}
\country{India}
}

\author{Prasenjit Karmakar}
\authornotemark[1]
\email{prasenjitkarmakar52282@gmail.com}
\affiliation{
    \department{Department of Computer Science \& Engineering}
    \institution{IIT Kharagpur}
    \country{India}
}

\author{Praveen Kumar Sharma}
\email{praveencs54@gmail.com}
\affiliation{
    \department{Department of Computer Science \& Engineering}
    \institution{NIT Durgapur}
    \country{India}
}

\author{Soumyajit Chatterjee}
\email{sjituit@gmail.com}
\affiliation{
    \department{Department of Computer Science \& Engineering}
    \institution{IIT Kharagpur}
    \country{India}
}

\author{Abhijit Roy}
\email{abhijitroy1998@gmail.com}
\affiliation{
    \department{Department of Computer Science \& Engineering}
    \institution{NIT Durgapur}
    \country{India}
}

\author{Santanu Mandal}
\email{santanucse16@gmail.com}
\affiliation{
    \department{Department of Computer Science \& Engineering}
    \institution{NIT Durgapur}
    \country{India}
}

\author{Subrata Nandi}
\email{subrata.nandi@gmail.com}
\affiliation{
    \department{Department of Computer Science \& Engineering}
    \institution{NIT Durgapur}
    \country{India}
}

\author{Sandip Chakraborty}
\email{sandipc@cse.iitkgp.ac.in}
\affiliation{
    \department{Department of Computer Science \& Engineering}
    \institution{IIT Kharagpur}
    \country{India}
}

\author{Mousumi Saha}
\email{msaha.nitd@gmail.com}
\affiliation{
    \department{Department of Computer Science \& Engineering}
    \institution{NIT Durgapur}
    \country{India}
}

\author{Sujoy Saha}
\email{sujoy.ju@gmail.com}
\affiliation{
    \department{Department of Computer Science \& Engineering}
    \institution{NIT Durgapur}
    \country{India}
}

\newcommand{\cmark}{\ding{51}}
\newcommand{\xmark}{\ding{55}}
\renewcommand{\shortauthors}{Pramanik, et al.}

\begin{abstract}
Efficient air quality sensing serves as one of the essential services provided in any recent smart city. Mostly facilitated by sparsely deployed Air Quality Monitoring Stations (AQMSs) that are difficult to install and maintain, the overall spatial variation heavily impacts air quality monitoring for locations far enough from these pre-deployed public infrastructures. To mitigate this, we in this paper propose a framework named \ourmethod{} that can annotate data obtained from a low-cost thermo-hygrometer (as the sole physical sensing device) with the AQI labels, with the help of additional publicly crawled Spatio-temporal information of that locality. At its core, \ourmethod{} exploits the temporal patterns from a set of readily available spatial features using an LSTM-based model and further enhances the overall quality of the annotation using temporal attention. From a thorough study of two different cities, we observe that \ourmethod{} can significantly help annotate the air quality data on a personal scale.
\end{abstract}



\maketitle

\section{Introduction}
Air pollution significantly impacts the daily life of individuals~\cite{world_health_organization_2018}, particularly the elderly residents of urban and metro cities. Various existing studies~\cite{krstic2011apparent,liu2018third,anjomshoaa2018challenges} indicate that air pollutants show significant spatial diversity across a city depending on different built environments like industrial establishments, housing complexes, parks, water bodies, city centers, transport depots, railway stations, and so on. Additionally, the distribution of pollutants also varies temporarily based on factors like weekends or weekdays, the weather condition of the day, special events such as Christmas or New Year, etc. City residents primarily follow the pollution reports that are publicly available to understand the air quality for a particular day. However, these pollution reports are typically captured from various \textit{Air Quality Monitoring Stations} (AQMS) deployed by Government agencies like Central Pollution Control Board (CPCB) or various private organizations at strategic locations like Airports, City Centers and within large industrial organizations establishments, etc. Therefore, the reports obtained from an AQMS primarily reflect the air quality of the site where the AQMS is deployed, which can be significantly different from the actual air quality of a location far from that site. 

Unfortunately, the number of AQMSs available in various urban and metro cities around the globe, particularly for developing and under-developed countries, is far less than the actual requirements~\cite{gu2018recurrent}. Consequently, city residents can hardly monitor the actual air quality of their localities and only get an approximate estimate from the publicly available reports~\cite{anjomshoaa2018challenges,kaya2020deep}. Therefore, it will be interesting to have a device or a system that can accurately estimate air quality at a personal level. The ensuing difficulty of deploying a proper infrastructure for measuring pollution levels can be attributed to several factors ranging from system deployment challenges to the financial challenges of installing and maintaining these systems. A typical environment sensing device contains specialized sensors that log and monitor pollutant levels. These sensors are not readily available and many times involve a critical process for calibrating their sensitivity. Additionally, the average cost of installing and maintaining an AQMS is also significantly high. Furthermore, due to high spatio-temporal variations and the inherent non-linearity of pollution samples, existing sparse AQMS infrastructure coverage is inadequate in providing fine-grained information even in the major cities. 

Air quality of a location is typically quantified using a metric called the \textit{Air Quality Index} (AQI) that combines measurements of various pollutants to give an indicator between $1$ (Good air quality) and $6$ (Hazardous). Interestingly, papers like~\cite{zheng2013u} have shown that more straightforward weather and meteorological features significantly correlate with the AQI of any locality. Although standard meteorological features like wind speed, wind gust, etc., do not change over a small area, these factors coupled with features like temperature, humidity, and spatial clutter can indeed help provide meaningful features for pre-training models targeted for AQI prediction~\cite{kleine2017modeling, zheng2013u}. Understanding these opportunities from existing literature, we define the primary goal of our paper as follows. \textit{Given the GPS along with readily available meteorological features like temperature and humidity from a low-cost thermo-hygrometer (THM), is it possible to develop a framework for a city-scale Air Quality Index (AQI) annotation?}

The primary challenge of designing a low-cost alternative for monitoring air quality at a personal scale is that the distribution of the target parameters like temperature and humidity depends on the climate and the demography of a city. Therefore, an AQI prediction model based on these parameters would be very city-specific, and we need to develop different pre-trained models for different target cities. Indeed, this is the major limitation of the existing models like~\cite{kleine2017modeling, zheng2013u,liu2018third,anjomshoaa2018challenges} that work only over a specific city or region. Developing a pre-trained model for each city is difficult, as the AQMSs that provide the training data for the model are costly devices. Consequently, in this paper, we start by developing a low-cost alternative of an AQMS, which is portable and can be deployed with minimum installation and management costs. We call these devices as \textit{Air Quality Monitoring Devices} (AQMDs).

While developing AQMDs, the primary challenge that we observe is that for any new device deployment, the device needs fine-grained calibration such that the device can augment and represent a real-time setup validated by existing standards of air quality monitoring. Recently, research works have explored various alternative air quality sensing modes like the use of portable devices~\cite{shindler2021development} and mobile/handheld devices~\cite{flow, airveda, aeroqual, airbeam}. In contrast to the existing AQMSs that use high-volume gravimetric sampling~\cite{measurement_of_pm10_particles_2002}, such alternatives use sensors like capacitive, resistive, electrochemical, optical, etc. However, the sensitivity of such sensors drifts with time, resulting in a lack of reliability of sampled data. Hence, for such modes, suitable calibration strategies need to be explored to obtain reliable samples. This makes the deployment and maintenance of such systems extremely challenging. 

Owing to these challenges, this paper develops a framework called \ourmethod{} that primarily has two objectives -- (1) utilize a thorough calibration method to periodically calibrate the AQMDs and then use the data from the AQMDs to develop a city-specific pre-trained model for AQI annotation, and (2) provide a real-time annotation module to predict and annotate the AQI level of a location using low-cost sensing. The first objective of \ourmethod{} augments the data sparsity problem. For the second objective, we observe that out of the various meteorological, temporal, and demographic parameters that impact the AQI of a city, temperature and humidity are very much location-specific and need on-spot measurements. In contrast, other parameters can be crawled from various publicly available sources. Consequently, we augment a simple, low-cost processing device having a THM along with a GPS to also report the AQI level of a location by utilizing the city-specific pre-trained models developed by \ourmethod{}.

\subsection{Our Contributions}
In comparison to the existing works, the contributions of this paper are as follows.\\

\noindent
\textbf{(1) Identifying the set of readily available features for AQI annotation:} The development of \ourmethod{} involves the selection of versatile features and modalities that are readily available with a known impact on the AQI of any place. This intelligent choice of features not only allows us to develop a generalized model but also ensures that during the deployment phase, localized sensing can be done using minimal hardware support.\\

\noindent
\textbf{(2) Creating a generalized pre-trained model for robust AQI annotation:} The crux of \ourmethod{} is the generalized, pre-trained model created using the data from pre-deployed AQMS(s) or AQMD(s) in a city. With data from these pre-deployed well-calibrated setups, the developed pre-trained model can generate labels for the localized sensing setups in other parts of the city. In contrast to the existing models that mostly use complex and computationally heavy learning techniques, we achieve comparable performance with a simple model by fine-tuning the system-specific setups, device calibration, and choosing a rich set of data from well-calibrated pre-deployed AQMS(s). More specifically, the model used in \ourmethod{} exploits the temporal patterns along with the readily available features and also utilizes the attention mechanism to understand specific temporal events for generating accurate AQI labels.\\

\noindent
\textbf{(3) Deployment and validation of \ourmethod{}:} We test \ourmethod{} over two different setups -- one by deploying four AQMDs within a $5$ sq km area of a metro city and another by crawling data from 12 AQMSs deployed within another metro city. We observe that for both the cities, \ourmethod{} can efficiently develop pre-trained models that can be used to annotate the AQI levels at different locations using a low-cost THM-based device with an F1-score of more than 60\%. Further, we observe that the prediction error of these models does not have a diverse impact. Therefore, city residents can use the model to have a personalized annotation of the air quality at their locations.\\

An initial version of this paper has been published in~\cite{sharma2021can}. In contrast to the previous version of the paper, we enriched the design of \ourmethod{} with a number of additional important features, particularly a number of temporal features like time of the day, seasons, month, day of the week, etc. In addition, we also developed an attention-based deep learning model for the robust prediction of AQI by enabling the model to search for the best combination of features depending on the temporal clustering of events. Finally, we extend the evaluation of \ourmethod{} by comparing the performance of different models and analyzing them under diverse scenarios. 

\subsection{Paper Organization}
The rest of the paper is organized as follows: Section~\ref{Survey} provides a detailed discussion of the related works, including cost-effective, ubiquitous air quality sensing, calibration of low-cost sensors, and AQI estimation. Next, Section~\ref{Data} highlights the challenges in developing and validating a low-cost AQMD that helps to gather ground truth information from a city with no or limited pre-deployed AQMS. In Section~\ref{dataset}, we analyze the in-house and publicly available datasets for two different cities in India and show the dependency of AQI on localized Temperature and Humidity. We further analyze the spatio-temporal impact on AQI distribution over individual monitoring devices. Section~\ref{frame} proposes our framework named \ourmethod{} that captures each step of the data processing pipeline following the development of city-specific pre-trained models for providing automated AQI annotation from the user's Thermo-hygrometer. Next, Section~\ref{Method} discusses the details on feature extraction and pre-training of city-specific models, followed by the evaluation of \ourmethod{} in Section~\ref{Eval}. Section~\ref{discussion} highlights some of the limitations of the \ourmethod{} framework. It further shows how deep learning solutions work well for relatively larger datasets and points out a few key areas of improvement over \ourmethod{} by leveraging Domain Adaptation techniques. Finally, Section~\ref{Conclude} concludes the paper.


\section{Related work}
\label{Survey}  
The critical condition of air across the major cities around the globe has led to a wide array of research in understanding the air quality with cost-effective methods, especially in a data-driven manner~\cite{zheng2013u, zheng2015forecasting, catlett2017array,chen2017open}. The primary focus so far has been to extend the air pollution measurement across the space (spatial measurements)~\cite{HOEK20087561} or over time (temporal measures)~\cite{zheng2015forecasting}. However, a large number of research studies have considered both space and time together (spatio-temporal measurements) to develop air quality prediction models~\cite{zheng2013u, catlett2017array,chen2017open}.

Most studies have focused on spatio-temporal modeling for AQI prediction because the pollution levels significantly vary over both space and time~\cite{tan2014characterizing, gu2018intracity}. This variability is due to the urban geographic factors~\cite{gulia2020sensor, HOEK20087561}, variation in traffic patterns~\cite{matte2013monitoring, zheng2013u}, among others. The federal agencies' precise sensing instruments (AQMS) cannot cover this variability as the density of sensors is very low and hence fails to capture the fine-grained variability of pollutants over large urban areas. Despite having 63 out of 100 most polluted cities in the world \footnote{https://www.ndtv.com/india-news/delhi-is-worlds-most-polluted-capital-for-2nd-straight-year-report-2836028 (Accessed: \today)}, India, according to CPCB's estimate, has only 4-8\% coverage of AQMSs\footnote{https://urbanemissions.info/ (Accessed: \today)}. As an alternative, the advent of low-cost sensing (LCS) (AQMD in our terminology) has increased the spatial and temporal coverage for AQI prediction~\cite{catlett2017array, chen2017open,liu2020low}. These sensors are often used in tandem with other available measurements to quantify air quality in locations where they are not placed. These measurements often use secondary features that impact air pollution like traffic count~\cite{matte2013monitoring}, pedestrian count~\cite{catlett2017array}, population density~\cite{zheng2013u}, land use/land cover~\cite{shukla2020mapping}, number of pollution sources or meteorological features like temperature, humidity, wind speed, wind direction, etc.~\cite{liu2020exploring}. This, combined with time-based patterns and the mobility of the sensors, can help understand the spatial dynamics of the air quality.

\subsection{Cost Effective Ubiquitous Air Quality Sensing}
Cost-effective air quality sensing encompasses various techniques to estimate the air quality either by the raw value of various pollutants or through the standardized AQI values. The initial line of research has focused on extending the AQMS measurements along with the local features~\cite{zheng2013u}. In~\cite{zheng2013u}, the authors have used reference-grade sensors and several local features that can be considered as secondary indicators for air pollution. However, in places where the density of AQMS is low, or features like human mobility, traffic count, etc., are not possible to retrieve, such methods are not applicable directly. Accordingly, a few works in the literature have focused on the development of portable yet robust air quality monitoring devices that can augment the existing AQMS~\cite{liu2020low,gulia2020sensor, purpleair-inc-2021, catlett2017array,chen2017open}. \tablename~\ref{tab:my-table} summarizes some of these works. 

\begin{table}[!htbp]
\caption{Survey of Systems (LCS: Low-cost Sensing)}
\centering
\resizebox{\textwidth}{!}{%
\begin{tabular}{|l|l|c|l|l|}
\hline
\textbf{Paper}                               
& \textbf{Type of Measurement}   
& \textbf{LCS Input} 
& \textbf{Physical Sensors} 
& \textbf{Area of Deployment}  \\ \hline
Purple Air\tablefootnote{https://www2.purpleair.com/ (Accessed: \today)}                                   
& PM2.5                          
& \cmark                
&  Plantower PMS5003                      
&  Worldwide    \\ \hline
Array of Things\cite{catlett2017array}                              
&  N$O_2$, $O_3$, Temp, Humidity 
& \cmark                
& AlphaSense                      
& Chicago     \\ \hline
Lj Chen\cite{chen2017open}                                      
& PM2.5                 
& \cmark                
& PlanTower                      
& Taiwan     \\ \hline
Koala Sensing \cite{liu2020low}                                
& PM2.5, CO                          
& \cmark                
& Plantower PMS1003                      
& Multi-Site     \\ \hline
Image Based Participatory Sensing\cite{liu2018third}            
& PM2.5                          
& \xmark                
& N/A                            
& Beijing     \\ \hline
Social Media Based Sensing \cite{pramanik2018aircalypse, pramanik2020aircalypse} 
& PM2.5                          
& \xmark                
& N/A                            
& New Delhi   \\ \hline
\end{tabular}%
}
\label{tab:my-table}
\end{table}

City-scale sensing of air pollutants has also been well-studied in the literature. The Array of Things project deployed in Chicago in 2018 has been state-of-the-art in sensing the city's health through multiple sensing modalities~\cite{catlett2017array}. In this project, the authors have deployed various sensors over $105$ sites across the city of Chicago to monitor several city-scale phenomena like urban heat islands, understanding the lake effect, etc. In~\cite{chen2017open}, the authors have performed a dense deployment of low-cost PM2.5 sensors across several places in Taiwan to create a participatory network of air quality particulate sensing. The primary goal of this work was to create a dense network of air quality sensors to monitor the air at a low cost. There have been other works like~\cite{liu2020low} that use low-cost sensing for measuring the $PM_{2.5}$ and carbon monoxide (CO), considering the calibration of only the CO sensor. However, their approach affects the reliability of the device. Mobile sensing through drones, named \textit{ Sensor-based Wireless Air Quality Monitoring Network} (SWAQMN), has been proposed by Gulia \textit{et. al.}~\cite{gulia2020sensor} to monitor real-time concentration of $PM_x$s. There are commercial devices such as `Purple Air'~\cite{purpleair-inc-2021} that sense $PM_{2.5}$ using pre-calibrated laser sensors, and each device contains two laser sensors of the same make and model to account for the inconsistencies. 

While the works discussed above are all related to primary sensing of the pollutants, other works indirectly measure air quality through various modalities like sound, social media, images, and other spatio-temporal factors~\cite{ghosh2018analyzing,pramanik2018aircalypse, pramanik2020aircalypse, liu2018third, sharma2021can}. In \cite{ghosh2018analyzing}, the authors have shown that acoustic signatures correlate well with air quality. Accordingly, they have developed a prediction model to estimate the air quality from the auditory noise information, especially for areas with a high density of traffic. Liu \textit{et al.}~\cite{liu2018third} have used images of the environment to measure the level of air quality. While this method is useful, the major disadvantage is extending it to places with not enough images labeled as the ground truth. Similarly, using social media posts, Pramanik \textit{et al.} have used various signals like influential users, public sentiment, and tweet volume to measure the level of air pollution in New Delhi, India~\cite{pramanik2018aircalypse,pramanik2020aircalypse}. However, with only 3-5 \% of tweets geotagged, effective mechanisms are required to locate the source of pollutants. These methods are still in the nascent stages and show accuracy even worse than low-cost sensing-based systems.

Therefore, we require a low-cost sensing-based localized AQI annotation platform by considering several primary and auxiliary air sensing modalities. A recent seminal work on the evaluation of low-cost sensing~\cite{peltier2021update} shows that the cost of maintaining the consistency and accuracy of these sensors is not low in the long run. Nevertheless, the report agrees that it is a valuable way of measuring particulate concentrations in moderate environments. In addition, there is ample scope for the refinement of models at a much granular level, especially for downstream tasks like the assessment of human exposure and dataset generation for the analysis of long-term trends once the device has been calibrated. Hence, developing a cost-effective low-cost sensor architecture that is scalable, reliable, and robust is a challenge.

 %


\begin{table}[!htbp]
\caption{Related Work on Software-based Calibration}
\scriptsize
\centering
\begin{tabular}{|l|l|l|l|l|l|}
\hline
\textbf{Ref} &
  \textbf{Sensors} &
  \textbf{Reference Monitors} &
  \textbf{Model} &
  \textbf{Features} \\ \hline
{\cite{lee2019efficient}} &
  AirBox by Edimax & 
  \begin{tabular}[c]{@{}l@{}} Taiwan Environmental Protection \\ Administration (TWEPA) Stations\end{tabular}&
  \begin{tabular}[c]{@{}l@{}}Generalized Additive \\ Model (GAM)\end{tabular} &
  PM2.5, T, RH \\ \hline
{\cite{desouza2022calibrating}} &
  \begin{tabular}[c]{@{}l@{}}PM2.5 (“Love My Air” \\ network, Denver)\end{tabular} &
  Federal Equivalent Monitor PM2.5 &
  \begin{tabular}[c]{@{}l@{}}21 ML algo each \\ with 4 correction \\ factors\end{tabular} &
  \begin{tabular}[c]{@{}l@{}}PM2.5, T, RH, \\ Dew, Time variant \\ factors\end{tabular} \\ \hline
{\cite{wang2019calibration}} &
  \begin{tabular}[c]{@{}l@{}}Particle monitor \\ (HK-B3, Hike, China)\end{tabular} &
  MicroPEM monitor (RTI, America) &
  Random Forest &
  PM2.5, T, RH \\ \hline
{\cite{cavaliere2018development}} &
  AIRQuino &
  TSI DustTrak &
  Linear Regression &
  PM2.5, PM10 \\ \hline
{\cite{zaidan2020intelligent}} &
  \begin{tabular}[c]{@{}l@{}}Clarity Corporation, \\ Berkeley, USA\end{tabular} &
  \begin{tabular}[c]{@{}l@{}} Monitor stations located at SMEAR\\ \& supersite Mäkelänkatu \end{tabular} &
  \begin{tabular}[c]{@{}l@{}}  Auto-Regressive \\  Model with LSTM \end{tabular}  &
  PM2.5 \\ \hline
{\cite{karaoghlanian2022low}} &
  PurpleAir PA-II-SD &
  Met One E-BAM PLUS &
  Linear Regression &
  PM2.5, PM10 \\ \hline
{\cite{chu2020spatial}} &
  \begin{tabular}[c]{@{}l@{}} AirBox, with PMS5003\\ optical particulate matter \\ sensor\end{tabular} &
  TWEPA’s Air Monitoring Network &
  Spatial Regression &
  PM2.5 \\ \hline
\end{tabular}
\label{tab:related_works_calibration}
\end{table}

\subsection{AQMD Calibration}
Due to the sensitivity drift in low-cost sensing, data reliability is a critical aspect that needs further analysis. Calibration can be of two types -- (i) \textit{software calibration}, also called soft calibration, and (ii) \textit{hardware calibration} or hard calibration. Hard calibration involves using specific volumetric calculations to measure that the sensors behave as they should~\cite{Alonso, Kularatna}. In~\cite{marathe2021currentsense}, the authors proposed \textit{CurrentSense}, a sensor fingerprint for detecting faults and drifts in environmental IoT sensors that could not be detected before without additional information. Their study used the concept that every electrical or electro-mechanical sensor draws current from the IoT device for its operation. By sampling the current drawn by the sensor, we can derive a unique electrical fingerprint that can distinguish between working, faulty, and malfunctioning sensors. The \textit{CurrentSense} fingerprint can also be used to monitor sensor health and improve sensor data quality. It is non-intrusive and can be applied to a wide variety of sensors. Moreover, this approach mostly focuses on detecting the faults and drifts, and after detection, the sensors must be brought to the lab for correction. In real practice, the monitoring network can be extensive and dynamic, so calibrating and maintaining such a monitoring network is cumbersome. While the advantage of this method is that the instrument itself is physically calibrated, The major drawback is the requirement of the sensors' physical presence in the hard calibration environment to calibrate it. In contrast, soft calibration involves software-based modifications to measure accurate data and can often be done remotely. A recent survey~\cite{concas2021low} focused on software-based calibration in low-cost sensors,  analyzed the effect of periodic re-calibration with the help of machine learning. We have summarized the types of soft calibration techniques in \tablename~\ref{tab:related_works_calibration} and described different models used for software-based calibration of the sensors.

In~\cite{lee2019efficient}, the authors have proposed a generalized additive (GAM)-based model to calibrate low-cost sensing by collecting data from regulatory stations in Taiwan. In a similar line, the authors in \cite{desouza2022calibrating} used $21$ different learning algorithms and developed four correction factors for each. They also deployed their sensing mechanism with a gold-standard reference monitor to obtain the calibration equation. The developed equation is then deployed to the deployed sensors to obtain the temporal and spatial trends of the network. In~\cite{wang2019calibration}, the authors have used a MicroPEM monitor (RTI, America) as a standard measurement device for particulate matters to calibrate the Hike monitors. The machine learning technique followed by 10-fold validation is used to obtain the concentration of particles. In a similar work~\cite{cavaliere2018development}, the authors have deployed low-cost air quality sensing devices in Florence near an official fixed air quality monitoring station and calibrated them. 

A more sophisticated calibration method has also been proposed in the literature. Zaidan \textit{et al.}~\cite{zaidan2020intelligent} have calibrated the meteorological parameters using the linear dynamic model, and the particulate matters are calibrated using non-linear models. In~\cite{karaoghlanian2022low}, the authors have used a calibration mechanism for the PurpleAir PA-II-SD that can measure the concentration of $PM_{2.5}$ and $PM_{10}$. They have used two high-fidelity Met One E-BAM PLUS placed at a single location in Beirut, Lebanon. The authors focused on the inter-sensor variability of PurpleAir sensors with their accuracy. They have used two linear regression models; the first model uses the entire concentration dataset, while the second model uses the 90\% quantile range to the concentration for better results without outliers. The authors consider spatially varying parameters in~\cite{chu2020spatial} by using low-cost sensing as well as regulatory stations. They performed regression analysis to explain the variability of the biases from the LCS. A summary of the sensors used, reference stations, and techniques with feature list by the above works is depicted in \tablename~\ref{tab:related_works_calibration}. Moreover, hardware calibration and sensitivity analysis are also crucial for improving the sensing reliability when deploying a system in any outdoor environment.

\begin{table*}
\scriptsize
\centering
\caption{Comparison of the above mentioned state of the art work and \ourmethod{}}
\label{tab_survey}
\resizebox{\columnwidth}{!}{
{\begin{tabular}{|l|l|l|c|c|c|l|c|} 
\hline
\multicolumn{1}{|c|}{\multirow{2}{*}{\textbf{Reference}}} & \multicolumn{1}{c|}{\multirow{2}{*}{\textbf{Objective}}} & \multicolumn{4}{c|}{\textbf{Input Feature Sources for Model Training}}   & \multicolumn{1}{c|}{\multirow{2}{*}{\textbf{ML Model}}} & \multirow{2}{*}{\textbf{\#AQI class}}  \\ 
\cline{3-6}
\multicolumn{1}{|c|}{}                                    & \multicolumn{1}{c|}{}                                    & \multicolumn{1}{c|}{\textbf{Meteorological}} & \textbf{Demographic}                & \textbf{Traffic Flow}             & \textbf{PM2.5/10}                       & \multicolumn{1}{c|}{}                                          &                                        \\ 
\hline
Zheng et. al.~\cite{zheng2013u}                            & AQI estimation                                           & Public Website                               & \multicolumn{1}{l|}{POI, road~data} & \multicolumn{1}{l|}{3000 Taxis}   & \xmark                                    & CRF  ANN         & 4                  \\ 
\hline
Lin Yijun et. al.~\cite{lin2018exploiting}                        & PM2.5 forecasting                                        & Dark Sky API                                 & \multicolumn{1}{l|}{OSM}            & \xmark                              & \multicolumn{1}{l|}{Public AQMS}        & CONV-RNN                                                                          & \xmark                                   \\ 
\hline
Xu x et. al.~\cite{xu2019multitask}                             & PM2.5 estimation                                         & Meteorological stations                      & \xmark                                & \xmark                              & \multicolumn{1}{l|}{Public AQMS}        & LSTM-Autoencoder                                    & \xmark                                   \\ 
\hline
Gu Ke et. al.~\cite{gu2018recurrent}                            & Air quality forecasting                                  & Self-deployed AQMSs                          & \xmark                                & \xmark                              & \multicolumn{1}{l|}{Self-deployed AQMS} & RAQP                 & \xmark                                   \\ 
\hline
Kaya et. al.~\cite{kaya2020deep}                             & PM10 forecasting                                         & Meteorological stations                      & \xmark                                & \multicolumn{1}{l|}{Traffic data} & \multicolumn{1}{l|}{Public AQMS}        & DFS         & \xmark                                   \\ 
\hline
Kleine et. al.~\cite{kleine2017modeling}                           & AQI estimation                                           & Meteorological stations                      & \xmark                                & \xmark                              & \xmark                                    & BT, L-SVM    & 3                                      \\ 
\hline
\ourmethod{}~\cite{sharma2021can}                                                 & AQI annotation                                           & THM, OpenWeather API                         & \multicolumn{1}{l|}{Gmaps API}      & \xmark                              & \xmark                                    & Random Forest  & 5                                      \\
\hline
This Paper                                                 & AQI annotation                                           & THM, OpenWeather API                         & \multicolumn{1}{l|}{Gmaps API}      & \xmark                              & \xmark                                    & LSTM with Attention  & 5                                      \\
\hline
\end{tabular}}
}
\end{table*}

\subsection{AQI-based Estimation} 
Several works have used machine learning-based spatio-temporal modeling to predict air pollution across a city. With the approaches either predicting the spatial distribution of pollutants via the AQI measurements or through forecasting based on historical measurements, the works have concentrated on predicting the future AQI of a locality based on the current measurements of pollutants. There are research works like~\cite{wu2019msstn,yi2018deep,han2018umeair,zheng2015forecasting,kok2017deep} which require a high volume of ground truth data and are particularly focused on temporal forecasting of pollutants rather than understanding the spatial distribution of the AQI over demography. Kaya \textit{et. al.}~\cite{kaya2020deep} have used meteorological, pollutant, and traffic data to forecast the concentration of $PM_{10}$ using a flexible deep model. In~\cite{lin2018exploiting}, the authors have proposed a diffusion convolution RNN model to forecast $PM_{2.5}$ particles for the next $24$ hours at a given location based on the meteorological and geographic data. In contrary, Zheng \textit{et. al.}~\cite{zheng2013u} have used a co-training approach to combine a spatial and a temporal classifier to estimate the AQI from various auxiliary data. In~\cite{xu2019multitask}, the authors have used a multi-layer LSTM and a stacked auto-encoder to estimate the pollution level from meteorological data, considering only local information, hence failing to capture the spatial spread of pollutants. Several works in the literature~\cite{kleine2017modeling,al2017prediction,qi2018deep,gu2018recurrent} have established the correlation among various meteorological data, like temperature, pressure, wind speed, and wind direction, among others, and proposed machine learning classifiers to predict the AQI. It is to be noted that the majority of these works demand continuous sensing of auxiliary information and assume uniformly available AQMS data across different city regions, which may not be available in most global cities. 
 
Our system \ourmethod{} has addressed the limitations of the existing works and provided an integrated framework ensuring the design of reliable sensing devices and developing suitable models for a city-wide fine-grained AQI annotation. \ourmethod{} combines sensing, calibration, and prediction to annotate temporal AQI measurements at a particular location. To make the system scalable, we develop and calibrate AQMDs to make the bootstrap pre-training easier for places where prior large-scale information is unavailable. Moreover, \ourmethod{}, apart from the GPS coordinates of the target location, requires minimal input from the user, viz., temperature and humidity, using a relatively low-cost device such as a simple THM (\textasciitilde 40 USD), compared to a direct sensing device like Purple Air (> 240 USD). A summary of the state-of-the-art methods is mentioned in \tablename~\ref{tab_survey} that compares \ourmethod{} with various existing works based on their objective, features source, ML model, and the number of AQI classes supported.



\section{Development of AQMD}
\label{Data}
\begin{figure}[!htbp]
    \captionsetup[subfigure]{}
    \begin{center}
        \subfloat[\label{dev_overview}]{
                \includegraphics[width=0.48\linewidth,keepaspectratio]{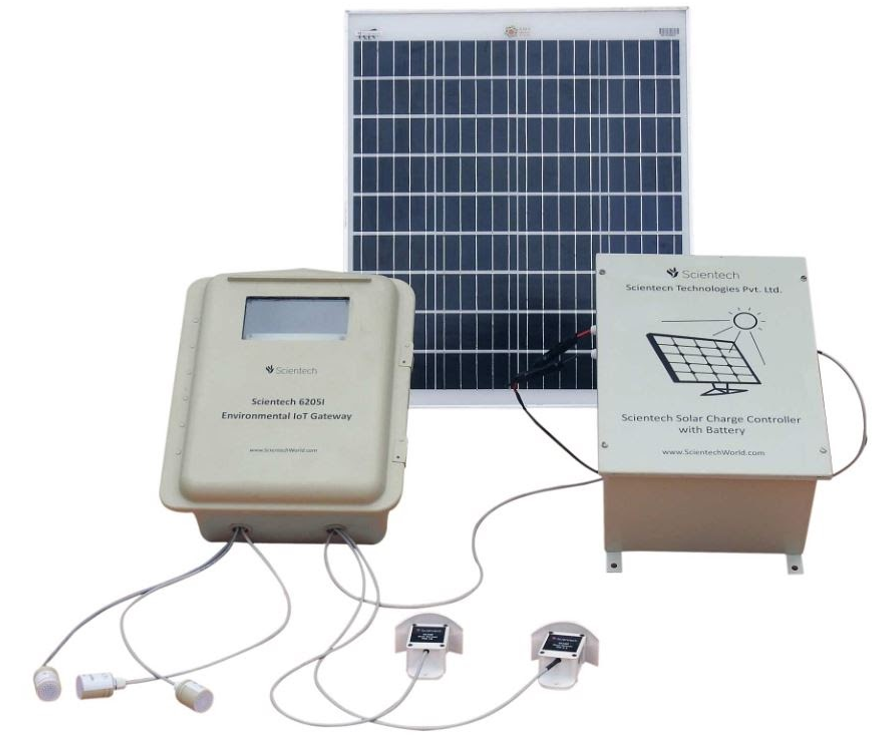}
        }
        \subfloat[\label{nit_device}]{
                \includegraphics[width=0.34\linewidth,keepaspectratio]{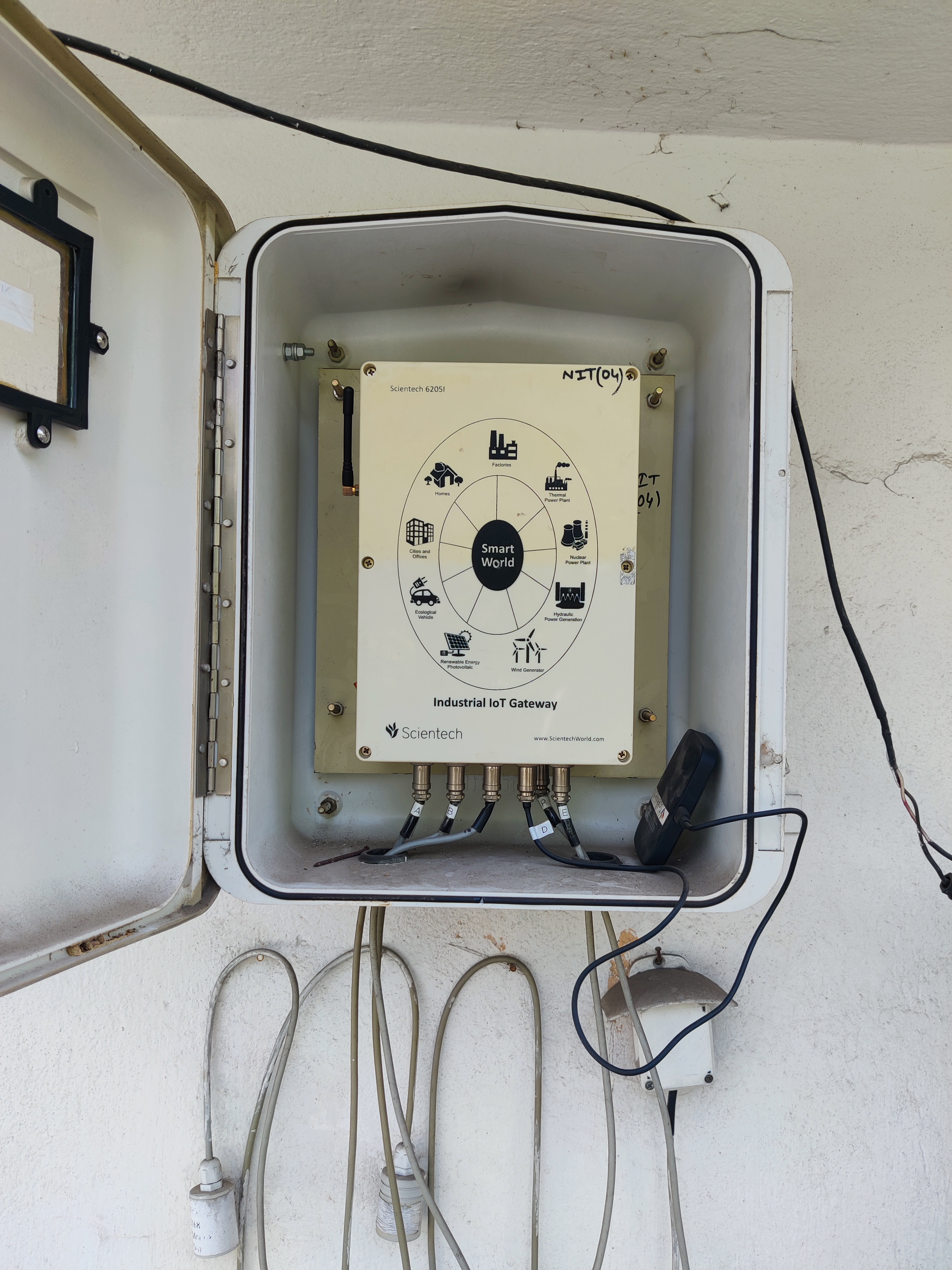}
        }
    \end{center}
    \caption{The AQMD device -(a) The system overview \& (b) The internal layout of a device deployment at one of the sites in \CityA{}}
    \label{aquamoho_deployment_nit}
\end{figure}

\begin{figure}
    \centering
    \subfloat[\label{aqms_1}]{
                \includegraphics[width=0.45\linewidth]{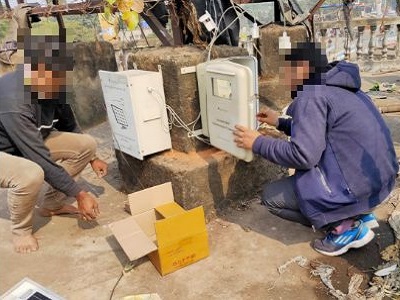}
     }
    \subfloat[\label{CPCB}]{
            \includegraphics[width=0.45\linewidth]{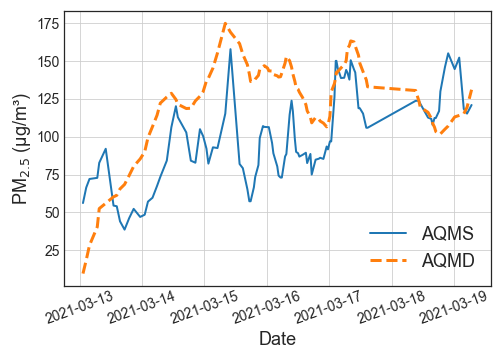}
     }
     \caption{(a) Deployment of the AQMDs in \CityA{}, (b) PM\textsubscript{2.5} concentration measured by AQMS, CPCB station and AQMD}
\end{figure}

The core idea of \ourmethod{} is to directly sense primary features like temperature \& humidity to find out AQI in a region given other spatio-temporal parameters that can be crawled from the public sources. However, for ground truth annotation, it is critical to develop low-cost AQMDs that can be deployed in strategic locations for sensing the concentration of air particles, pollutants, and meteorological features. It can be noted that these AQMDs increase the range and accuracy of air quality monitoring by augmenting the AQMSs deployed in a city. However, \ourmethod{} works as long as the AQI values from some nearby strategic locations are available, either from other AQMSs or from AQMDs. Thus, the developed AQMDs help us get the ground-truth AQI values to evaluate the performance of \ourmethod{} AQI prediction. \figurename{ \ref{aqms_1}} \& \figurename{ \ref{aquamoho_deployment_nit}} show the deployment of an AQMD at a strategic location in the \CityA{}. 

\begin{table}[!htbp]
\centering
\caption{System Specifications of Air Quality Monitoring Device (AQMD)}
\label{tab:sys_spec_aqmd}
\resizebox{\textwidth}{!}{%
\begin{tabular}{|ll|l|l|cclc|}
\cline{1-2} \cline{4-8}
\multicolumn{2}{|c|}{\textbf{System Architecture}} &
  \multicolumn{1}{c|}{\textbf{}} &
  \multicolumn{1}{c|}{\textbf{Sensor Name}} &
  \multicolumn{4}{c|}{\textbf{Sensor Operation Details}} \\ \cline{1-2} \cline{4-8} 
\multicolumn{1}{|c|}{\textbf{\begin{tabular}[c]{@{}l@{}}Processor Board\\  Details\end{tabular}}} &
  \multicolumn{1}{l|}{\begin{tabular}[c]{@{}l@{}}64bit ARMv7 Quad \\ Core Processor 1.2GHz\end{tabular}} &
  \multicolumn{1}{c|}{\textbf{}} &
  \multicolumn{1}{c|}{\textbf{}} &
  \multicolumn{1}{c|}{\textbf{Range}} &
  \multicolumn{1}{c|}{\textbf{Response Time}} &
  \multicolumn{1}{c|}{\textbf{Operational Range}} &
  \textbf{Remarks} \\ \cline{1-2} \cline{4-8} 
\multicolumn{1}{|l|}{\textbf{Memory}} &
  \cellcolor[HTML]{FFFFFF}\begin{tabular}[c]{@{}l@{}}1GB RAM; \\ 32GB Internal Memory\end{tabular} &
   &
  \textbf{\begin{tabular}[c]{@{}l@{}}Temperature \&\\ Humidity\end{tabular}} &
  \multicolumn{1}{c|}{\begin{tabular}[c]{@{}c@{}}0- 100 Degree C/\\ 0-95\% RH\end{tabular}} &
  \multicolumn{1}{c|}{1 Sec.} &
  \multicolumn{1}{l|}{N/A} &
  \textbf{-} \\ \cline{1-2} \cline{4-8} 
\multicolumn{1}{|l|}{\textbf{\begin{tabular}[c]{@{}l@{}}Network \\ Connectivity\end{tabular}}} &
  \begin{tabular}[c]{@{}l@{}}Wi-Fi, Ethernet \&\\  GSM Modem\end{tabular} &
   &
  \textbf{\begin{tabular}[c]{@{}l@{}}Dust Sensor \\ (PM2.5/PM10)\end{tabular}} &
  \multicolumn{1}{c|}{0-1000ug/m3} &
  \multicolumn{1}{c|}{10 Sec.} &
  \multicolumn{1}{l|}{\begin{tabular}[c]{@{}l@{}}30$\sim$70 Degree/\\ 15$\sim$90\%RH\end{tabular}} &
  \textbf{-} \\ \cline{1-2} \cline{4-8} 
\multicolumn{1}{|l|}{\textbf{Scan Rate}} &
  1 Minute &
   &
  \textbf{CO2} &
  \multicolumn{1}{c|}{300-4000ppm} &
  \multicolumn{1}{c|}{120 Sec.} &
  \multicolumn{1}{l|}{\begin{tabular}[c]{@{}l@{}}5$\sim$50 C Degree/\\ 10$\sim$90\%RH/\\ 811hPa$\sim$1216hPa\end{tabular}} &
  \multicolumn{1}{l|}{Resolution: 10 ppm} \\ \cline{1-2} \cline{4-8} 
\multicolumn{1}{|l|}{\textbf{Power Supply}} &
  \begin{tabular}[c]{@{}l@{}}Battery operated (12V-12Ah) \\  with Solar Panel (12V-50W) \end{tabular} &
   &
  \textbf{NO2} &
  \multicolumn{1}{c|}{0-20ppm} &
  \multicolumn{1}{c|}{30 Sec.} &
  \multicolumn{1}{l|}{\begin{tabular}[c]{@{}l@{}}0$\sim$50 Degree/\\ 15$\sim$90\%RH/\\ 811hPa$\sim$1216hPa\end{tabular}} &
  \multicolumn{1}{l|}{Resolution: 0.1ppm} \\ \cline{1-2} \cline{4-8} 
\end{tabular}%
}
\end{table}

AQMDs use low-cost electronic sensors in place of the high-volume gravimetric sensors that are used in federal-grade AQMSs. Therefore, these devices are calibrated to achieve a level of reliability for deployment in a real scenario. We then prepared the dataset by deploying four AQMDs in a target city (\textit{\CityA{}}\footnote{This is a major industrial Tier-II city in India, having an area of $154$sqkm, with a population density of $3700$/sqkm. Due to the heavy industry establishments (having steel plants, thermal power stations, and chemical industries, among others), the pollution levels in different parts of the city fluctuate at different times. Therefore, we considered this city as an ideal case for our study.}) at four different locations. The distribution of the collected data is then analyzed to get a clear insight into the air quality across the city. We have also selected a major city in India and collected the data through publicly available data sources to evaluate \ourmethod{}. Therefore, according to the experimental locations, we defined the data in two forms, \textit{in-house data} which are collected through the developed AQMDs, and \textit{publicly available data} which are crawled from publicly available sources.

\subsection{Device Development and Calibration}
AQMD has been developed with the goal of reliable sensing at a low cost. To that end, in partnership with an original manufacturer, we designed the AQMD for deployment in \CityA{}, which has only one AQMS and, therefore, is unable to sense the city with sufficient granularity. The system specifications of the AQMD are given in \tablename~\ref{tab:sys_spec_aqmd}. The device components are organized in three physical layers -- the \textbf{sensing layer}: top layer where all the sensors are present, the \textbf{controller layer}: where the microcontroller is placed, and the \textbf{power layer}: which deals with the power supply to the microcontroller and the sensors. For the sensing layer, the connected sensors are the Dust Particle sensor, Temperature Humidity sensor, NO\textsubscript{2} sensor, and CO\textsubscript{2} sensor. \tablename~\ref{tab:sys_spec_aqmd} summarizes the performance characteristics of the sensors. For the controller layer, the device contains an ARM v7-based Single Board Computer, which is robust and can frequently poll from the connected sensors, as shown in \tablename~\ref{tab:sys_spec_aqmd}. The device consists of local storage in addition to cloud storage for storing the data. It has the network capability to transmit via Wi-Fi as well as 4G through the MQTT protocol. We utilize the 4G module for connectivity. Finally, the power layer contains the module for solar to utilize the solar power (12V-50W) and has a power storage of 12V-12Ah. \figurename{ \ref{aquamoho_deployment_nit}} shows the outdoor setup of the device along with the different components.

Among the different pollutant and particle data collected from the AQMDs, we primarily analyze the $PM_{2.5}$, as many existing studies~\cite{cheng2016status,kleine2017modeling} confirm that $PM_{2.5}$ concentration is the most dominating pollution factor in an outdoor scenario. AQMDs utilize the sensor to measure $PM_{2.5}$ and two meteorological parameters: {\it temperature}, and {\it relative humidity}, with different sensors, viz., Metal Oxide Semiconductor (MOS), Optical, and Capacitive sensors. As these AQMDs use low-cost sensors susceptible to erroneous sensing, so we use a thorough validation and calibration mechanism. The details follow. 

The sensors used in AQMDs are susceptible to erroneous measurement due to the shifting of zero or baseline value and due to the electronic aging of the components, also known as \textit{baseline drift}. Initially, we validated the sensors to resolve this issue as follows. The devices are kept inside a vacuum chamber where $N_2$ gas is purged inside to make it vacuum. In such a condition, the sensor readings are supposed to reach zero readings. The observed values are then marked as their respective baselines. This process is also called the \textit{zero-air calibration}. Then we validated the sensors in two phases, \textit{pre-deployment validation:} to validate the precision and sensitivity before deployment, and \textit{post-deployment validation:} to validate the accuracy after deploying it at different locations.

\subsubsection{Pre-Deployment Validation} This consists of validating the AQMD on a lab scale under multiple conditions to assess their precision as well as sensitivity. The details are as follows.\\

\noindent\textit{External Reference Validation:}
In this method, the sensors have been validated using external references. The AQMS placed by the Central Pollution Control Board (CPCB) situated at \CityA{} is taken as a reference. Our AQMD has been placed at a distance of $\approx 300$ meters from the reference AQMS. The data collected by both sources have been compared. As noted from \figurename{ \ref{CPCB}}, we observe that our AQMD shows a similar behavior when compared with the AQMS. The other custom AQMDs are then validated using the already validated AQMD. Let $\mathcal{A}_v$ be the AQMD that has been validated with a federal AQMS. In the following steps, we validate other AQMDs $\mathcal{A}_n$ with the help of $\mathcal{A}_v$ using the methods as discussed next.\\

\begin{figure}
    \centering
     \includegraphics[width=0.5\columnwidth]{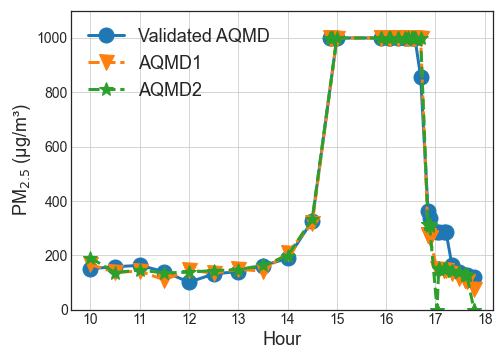}
         \caption{PM\textsubscript{2.5} concentration measured by different AQMDs in a controlled fire event}
         \label{event}
\end{figure}

\noindent\textit{Precision Assurance: (In normal condition Indoors)}
Precision is the degree of reproducibility, i.e., if the same value is sensed each time under the same environment. We have taken the measurements in the same environment to test the reproducibility. In normal conditions, the AQMDs are placed inside the lab, and on analyzing the collected data, it shows the similar behavior of the devices as shown in \figurename{ \ref{event}}. The figure shows identical variations, and the p-value of $0.30$ of the hypothesis test (P-test~\cite{rice2006mathematical}) also supports that the AQMDs have similar behavior.\\

\noindent\textit{Sensitivity Analysis: (Indoors with an event)} The sensitivity of a sensor is defined as the changes in the measurement of changing the environment. The basic idea of such calibration is to trigger an external event that results in a sudden fluctuation in the pollution level. In a natural setup, the devices should be able to capture such fluctuations correctly. To generate such external events, we first placed $\mathcal{A}_v$ and $\mathcal{A}_n$ in a room under the same environment. We then generated the event by lighting a fire, which caused smoke in the room. Due to the accumulation of smoke inside the room, $PM_{2.5}$ concentration increases sharply, which can be seen in the \figurename{ \ref{event}}. The event's effect decreases on opening the doors and windows, i.e., by creating proper ventilation. The rise and drop in the particle concentration confirm the sensitivity when the measured values indicate similar patterns for $\mathcal{A}_v$ and $\mathcal{A}_n$. 

\subsubsection{Post-Deployment Validation} 
After deployment, the AQMDs are validated through the accuracy analysis concerning the already validated AQMD $\mathcal{A}_v$. The accuracy of a system is the degree of closeness of the measured value to that of the system's true value \footnote{\url{https://en.wikipedia.org/wiki/Accuracy\_and\_precision} (Accessed: \today)}. We have used the pre-validated AQMD $\mathcal{A}_v$ and placed it with each deployed AQMD $\mathcal{A}_n$ for a specific duration and used the collected data to analyze the accuracy. We have analyzed the similarity using hypothesis testing (P-test) that results in a p-value of more than $0.40$ for all the AQMDs supporting the null hypothesis that there is no significant difference between $\mathcal{A}_v$ and $\mathcal{A}_n$.

\subsection{System Deployment}
\begin{figure}
 \centering 
 \subfloat[\label{city1_devs}]{
             \includegraphics[width=0.37\linewidth]{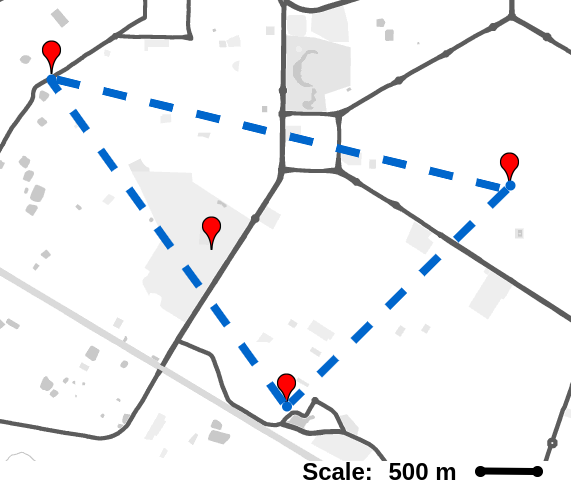}
   }
   \subfloat[\label{city2_devs}]{
               \includegraphics[width=0.37\linewidth]{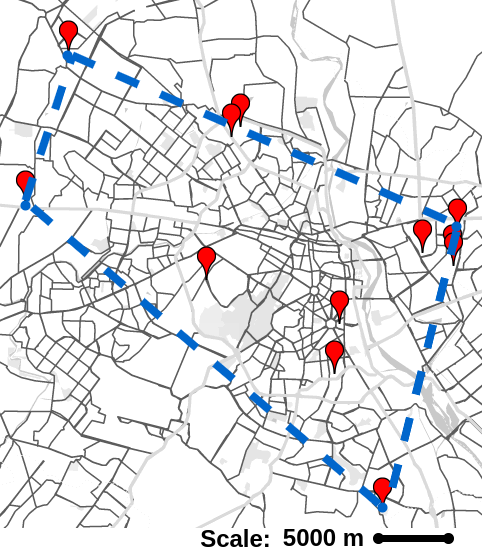}
   }
    \caption{The red dots and blue bounded figures represents the position of AQMDs and area of coverage respectively in (a) \CityA{}  \& (b) \CityB{}}
    \label{city_devs}
\end{figure}

We have selected four locations in \CityA{} as depicted in \figurename{ \ref{city1_devs}} and deployed four AQMDs, as shown in \figurename{ \ref{aqms_1}}, for sensing the pollutants, particles, and meteorological parameters. The AQMDs provide the data samples with an adaptive sampling rate, i.e., they provide the samples whenever the new data sample differs from the previous data sample and store the data in the cloud. The Geospatial imagery of the locations is used to extract the information regarding the \textit{Point of Interests} (PoI) at different locations such as Natural land, Park, Road networks, Educational institutions, etc. \tablename~\ref{TabAQMS1} explains the intuition behind the selection of locations to deploy the AQMDs. AQMD-1 is placed in a residential zone (0.57 sq km) with a natural land cover of $72\%$. AQMD-2 has been deployed in a region with three bus depots and crowded marketplaces. AQMD-3 is deployed at a location in the center of the city and consists of almost all the PoIs. One thermal power station is also situated near the site where AQMD-3 is deployed. AQMD-4 is deployed at a location with an educational institution having a large green area. We have analyzed the basic demography of the city to deploy the AQMDs to ensure that they can cover different behaviors of the city in terms of pollution exposure to the public. To get a closer estimate of the AQI at a location, this is required.

\begin{table}
\scriptsize
\centering 
\caption{Choice of locations to deploy the AQMDs in \CityA{}. Here, a location is considered a virtual square around the location with each side of length 1 km.}
\label{TabAQMS1}
\begin{tabular}{p{1.2 cm}p{6.8 cm}}
\toprule
Device ID     & Remarks 
\\ \midrule
AQMD-1 & This location is a residential zone with small educational regions, high natural land coverage of 71.42\%, and a road network coverage of 24\%.  
\\
AQMD-2 & This location is a densely populated market area that has multiple bus depots. The region also has natural land and human-made structures coverage of 48\% and 17\%, respectively. There are many road networks in the area, almost covering 37\% of the region. 
\\
AQMD-3 & This location can be regarded as the heart of the city. It consists of PoIs like various attractions, including shopping complexes, food plazas, and cinema halls with hospitals. The location is situated near industrial areas and has a high presence of road networks, including highways. 
\\
AQMD-4 & This location has the highest presence of educational institutions, which occupies 22\% of this location. Here, abundant greenery is also present, with 39\% natural land coverage and 12\% human-made structures.
\\ \bottomrule
\end{tabular}
\end{table}

\subsection{Publicly Available Data}
We have selected a city in India, \CityB{}, where 37 AQMSs are deployed sparsely in the city. We have collected the data samples of air quality with few other meteorological data through a publicly available web portal\footnote{\url{https://app.cpcbccr.com/ccr/\#/} (Accessed: \today)} provided by Central Pollution Control Board (CPCB), and Indian Meteorological Data (IMD). The portal provides different environmental parameters such as $PM_{2.5}$, $PM_{10}$, $NO$, $NO_2$, $NO_x$, $SO_2$, $CO$, etc. We are interested in parameters that impact the most and can be used as features in the prediction model. We have selected a set of parameters such as $PM_{2.5}$, temperature, relative humidity, wind speed, wind direction, barometric pressure, and atmospheric temperature. We have crawled the data for the last $17$ months in different chunks ($3$ months) for all the AQMSs in \CityB{} ($37$ stations). However, we found that only $12$ out of the $37$ stations have the required data; while most of them do not provide all the metrics we are interested in, some have missing data over a long time duration in multiple instances. Therefore, $12$ AQMSs in \CityB{} can be used for evaluating our proposed methodology, and their locations have been depicted in \figurename{ \ref{city2_devs}}.

In the next section, we look into the datasets in hand to analyze the different parameters and identify the relevant ones for further use.

\section{Preliminary Study of the Data }
\label{dataset}
For designing \ourmethod{}, we first analyze the collected data to explore a few insights about the spatio-temporal patterns of AQI distribution along with the impact of various spatial and temporal parameters on the measured AQI values. One of the prime cruxes behind our design is that the spatial features for a location can be extracted from the publicly available topographical and GIS information, which can be clubbed with the temporal meteorological features to predict the AQI level of a location. For this purpose, the data collected from \CityA{} and \CityB{} are preprocessed, followed by a thorough study of its distribution based on the AQI classes both spatially and temporally. The details are as follows.

\subsection{Asynchronous Sensor Polling}
\ourmethod{} describes the deployment of four AQMDs in \CityA{}; these AQMDs are designed in a way that enforces it to update the data whenever a new value is sensed. This process reduces the system overhead to generate the samples in each trigger, improving system performance and preventing data duplication. As a result, the data sampling becomes irregular, and the collected data needs to be preprocessed to obtain a fixed sampling rate. We simply preprocess the data by replacing the missing values using the existing values from the previous sampling window. However, regarding the data collected for \CityB{}, we perform no explicit preprocessing. We analyze the distribution of AQI based on $PM_{2.5}$ concentration to get an overview of the target location's air quality. The AQI distribution helps us understand the need for further processing to obtain the required results through predictive modeling. We have considered 5 AQI classes - AQI 1 (0-30, Good), AQI 2 (31-60, Satisfactory), AQI 3 (61-90, Moderately Polluted), AQI 4 (91-120, Poor), \& AQI 5 (121-250, Very Poor). We ignore AQI 6 as we obtain a very less sample for it for both cities. Based on this preprocessing of the data, we next analyze its insights. 

 \begin{figure}
     \captionsetup[subfigure]{}
     	\begin{center}
                 \subfloat[\label{city1_sp_AQI}]{
                 			\includegraphics[width=0.45\linewidth,keepaspectratio]{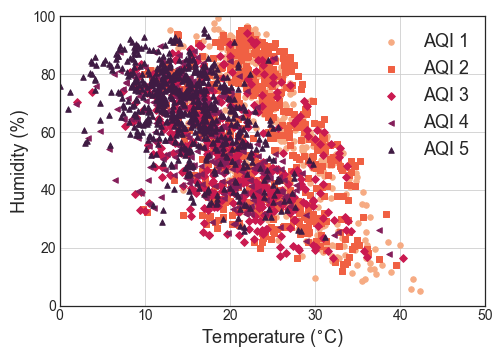}
                 }%
                 \subfloat[\label{city2_sp_AQI}]{
                 			\includegraphics[width=0.45\linewidth,keepaspectratio]{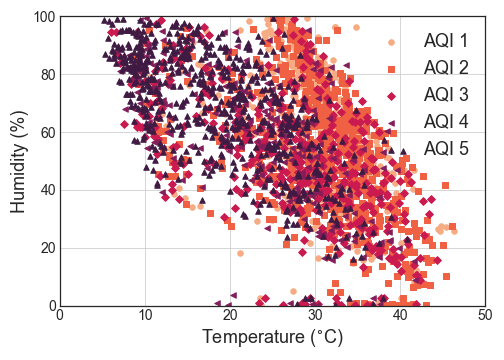}
                 }
         \end{center}
     \caption{Correlation of THM based features with the AQI classes in (a) \CityA{} \& (b) \CityB{}.}
     \label{th_aqi_sp}
\end{figure}
\subsection{Thermo-hygrometer(THM)-based Features vs AQI} 
We start by analyzing the temperature and humidity, which can be collected from a readily available THM. We analyze the correlation between these collected factors with AQI classes. \figurename{ \ref{city1_sp_AQI}} and \figurename{ \ref{city2_sp_AQI}} show the strong correlation between the temperature and humidity with different AQI classes for both the cities. Furthermore, a deep observation from the figure shows that the low humidity and high temperature correspond to the good AQI classes. Increasing the humidity with an increase in temperature shows the deterioration of air quality toward severe AQI. Additionally, the observations are persistent across the cities. Hence, the temperature and humidity of a location in a city are potential parameters for generating the AQI annotation for that location. 

Interestingly, \figurename{ \ref{th_aqi_sp}} indicates that the exact AQI distribution concerning the two meteorological parameters is indeed different for the two cities, although the pattern remains the same. For example, we observe from \figurename{ \ref{city1_sp_AQI}} that high AQIs are more dense near 60-80\% humidity and 5$^\circ$-20$^\circ$C temperature in \CityA{}. Whereas, \figurename{ \ref{city2_sp_AQI}} indicates that the high AQIs are spread within 60-100\% humidity and 0$^\circ$-25$^\circ$C temperature for \CityB{}. These differences in the spatial spread of the AQI values are primarily due to the climate of a city; for example, some city (like \CityB{}) frequently observes $<5^\circ$C temperature during winter, whereas such a low temperature is rare for some cities (like \CityA{}). This difference in the climate governs the impact pattern of meteorological parameters on the AQI values, which vary across cities. 

In the following subsections, we discuss the impact of the spatial and temporal factors on AQI distribution from the four deployed AQMDs in \CityA{} and twelve publicly deployed AQMSs in \CityB{}. For this purpose, we analyze the data collected for $12$ months and $17$ months, respectively, for the two cities.

\subsection{Impact of Spatial Parameters on AQI Distribution}
\begin{figure}
    \captionsetup[subfigure]{}
    	\begin{center}
                \subfloat[\label{dgp_dist}]{
                			\includegraphics[width=0.45\linewidth,keepaspectratio]{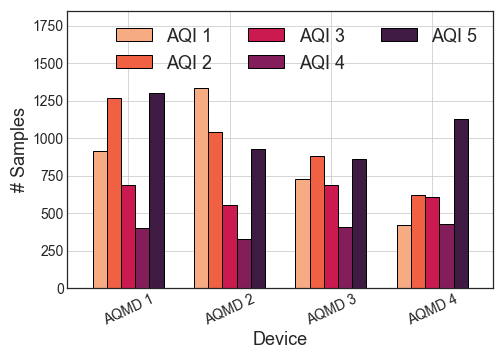}
                }%
                \subfloat[\label{delhi_dist}]{
                			\includegraphics[width=0.45\linewidth,keepaspectratio]{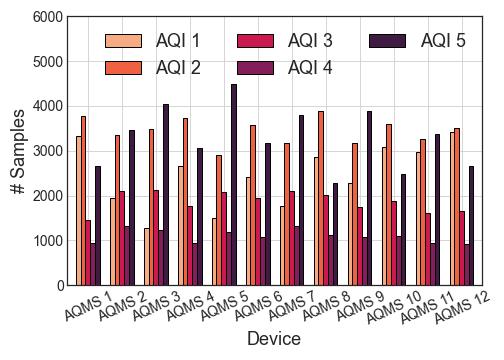}
                }
        \end{center}
     \caption{Device-wise AQI distribution of (a) \CityA{} \& (b) \CityB{}.}
     \label{Distrib_city}
\end{figure}

We first analyze the impact of various spatial features on the AQI distribution. As shown in \figurename{ \ref{Distrib_city}}, we observe that for both cities, there is indeed an impact of change in the location of the devices on AQI distribution. For example, in \CityA{}, all the AQMDs show similar behavior for all the AQI classes apart from the distribution of AQI classes 1 and 2, which fluctuate due to the city's heterogeneity. Similar behavior can be observed in \CityB{} as well. Therefore, looking into the spatial parameters can help us in better AQI estimation. 

\begin{figure}
    \captionsetup[subfigure]{}
    	\begin{center}
                \subfloat[\label{dgp_tz_dist}]{
                			\includegraphics[width=0.45\linewidth,keepaspectratio]{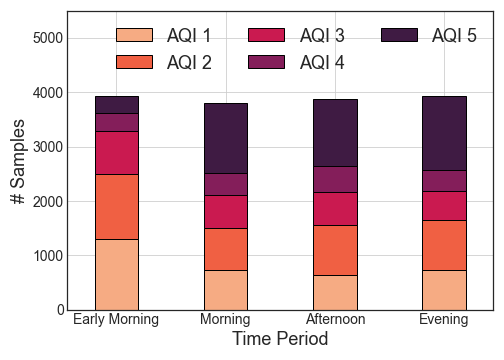}
                }%
                \subfloat[\label{delhi_tz_dist}]{
                			\includegraphics[width=0.45\linewidth,keepaspectratio]{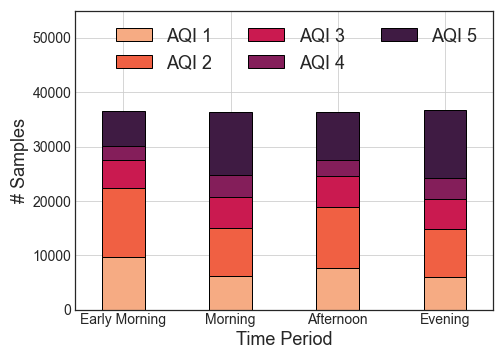}
                }
        \end{center}
     \caption{AQI distribution of (a) \CityA{} \& (b) \CityB{} at different times of the day. The time distributions are uniformly selected and are of 6 hours, starting from 00:00 Hrs.}
     \label{Distrib_city_tz}
\end{figure}

\subsection{Impact of Temporal Features on AQI Distribution}
\label{subsection:temporal}
Building on top of our prior work \cite{sharma2021can}, we analyze the temporal aspect of the AQI distribution using the collected data from the four deployed AQMDs in \CityA{} and 12 publicly deployed AQMSs in \CityB{}, for $12$ months and $17$ months, respectively. Our primary observation here, as shown in \figurename{ \ref{Distrib_city_tz}}, is that for both cities, based on time, the distribution of AQIs varies significantly. Interestingly, we observe similar behavior in both cities. \CityA{} and \CityB{} exhibit their highest levels of air pollution during the morning and the evening hours. In contrast, the concentration of air pollutants is moderate during the afternoon and the lowest in the early morning hours. However, the percentage of AQI-5 instances during the early morning is significantly higher in \CityB{} when compared to \CityA{}. Therefore, a robust AQI annotation model should consider such temporal variations to predict localized AQI values accurately. 

\subsection{Localized Temperature and Humidity} 
\begin{figure}
    \captionsetup[subfigure]{}
    	\begin{center}
                \subfloat[\label{dgp_dT_dH}]{
                			\includegraphics[width=0.45\linewidth,keepaspectratio]{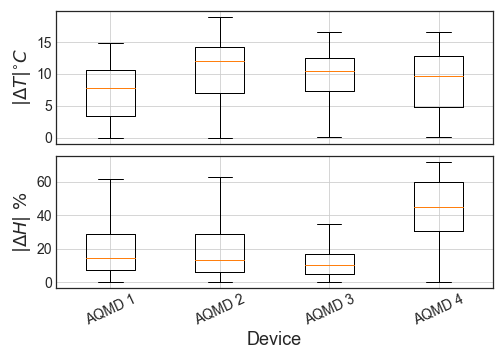}
                }%
                \subfloat[\label{delhi_dT_dH}]{
                			\includegraphics[width=0.45\linewidth,keepaspectratio]{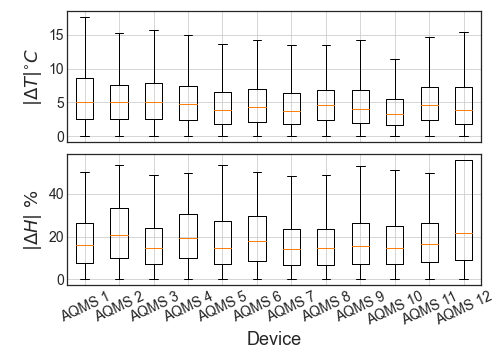}
                }
        \end{center}
     \caption{Device wise temperature and humidity difference with citywide median temperature and humidity of (a) \CityA{} \& (b) \CityB{}.}
     \label{dT_dH_city}
\end{figure}

Here, to justify the importance of localized temperature and humidity for the annotation of AQI values, we have analyzed the citywide temperature and humidity measurements collected through the open-sourced web API. We have compared the citywide median temperature and humidity with the observed temperature and humidity values measured at different AQMDs for \CityA{}. A similar comparison is also made for the AQMSs in \CityB{}. The deviation in the citywide median temperature and humidity of \CityA{} and \CityB{}, for the AQMDs/AQMSs, is shown in \figurename{ \ref{dT_dH_city}}. The figure shows that the deviation is significantly high, which infers an uneven distribution of temperature and humidity in a city. Hence, the localization of temperature and humidity sensing is crucial for achieving good annotation accuracy.

\subsection{Key Takeaways}
These analyses found that temperature and humidity are crucial parameters that widely vary across different city locations, and the air quality heavily depends on these parameters. Spatial factors like the presence of greenery or urbanization also impact air quality. In addition, temporal factors like the time of the day play a crucial role in determining the AQI values. As we observe, the air quality depends on the temporal urban characteristics. Certain situations like the restriction of goods vehicles during the day times, the usual rush before and after office hours, etc., impact the localized air quality in a city. Interestingly, meteorological parameters like temperature and humidity also capture the effect of spatial topography. For example, the average temperature near an industrial factory is likely to be more than the temperature at a park. Consequently, these two features become the decisive factors in determining the localized air quality of a city. However, the level of impact varies across cities; therefore, a pre-trained model developed for one city is not directly transferable to another city. Considering these factors, we develop a robust model for localized AQI annotation based on thermo-hygrometer sensing, as discussed in the next section. 

\section{System Overview}
\label{frame}
\ourmethod{} exploits the above observations and develops a low-cost system for annotating the indicative air quality of a location over a personal device. For this purpose, we first develop city-specific pre-trained models by utilizing various parameters obtained through AQMDs/AQMSs. Then during real-time queries, a THM provides the temperature and humidity of the queried location, and the system crawls the spatial features from publicly available web-based resources. It finally estimates the indicative air quality of that location by utilizing the measured parameters and the city-specific pre-trained model.
\begin{figure}
  \centering
  \includegraphics[width=0.90\linewidth]{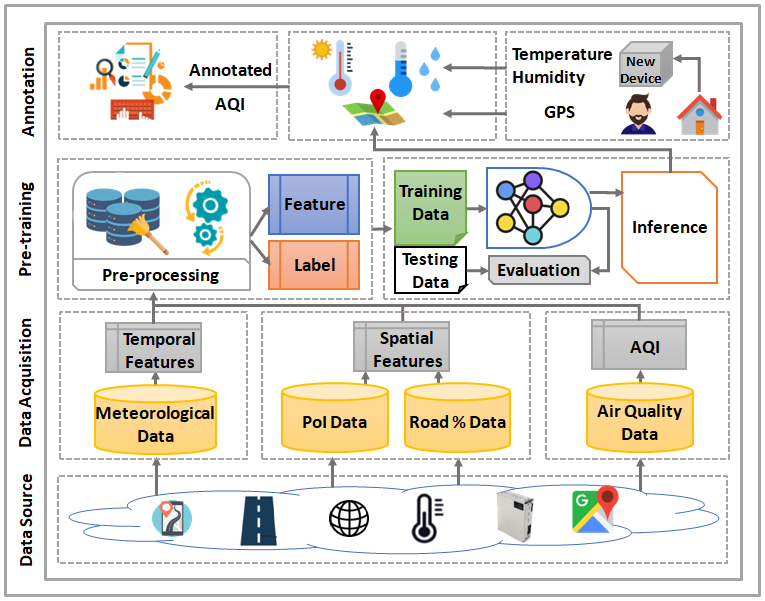}
  \caption{Broad overview of \ourmethod{} Framework.}
  \label{framework1}
\end{figure}

\figurename{~\ref{framework1}} shows a broad overview of \ourmethod{} framework. The system contains four primary modules. The first module includes the delineation of the data sources. These data sources contain the AQMDs, public AQMSs, public GIS information, road network information, etc., and are used to develop the pre-trained city-specific model. Subsequently, the second module extracts generic features from the available sensors, open-source GIS, and weather information-related APIs.
Additionally, in this step, the framework also considers the input from dust sensors available from the deployed AQMDs to measure $PM_{2.5}$ and further compute the AQI information. Together with the AQI information, all these features form the training data for that particular city. Subsequently, this training data is adequately pre-processed in the third module and then used to create pre-trained models. Finally, the last module serves as the data annotation end-point for any user who queries the system with information obtained from a low-cost THM and their location details (GPS coordinates) and obtains the AQI annotations. This finally annotated dataset acts as an alternate source of AQI information for the low-cost device without the explicit requirement of specialized sensing for $PM_{2.5}$. As a consequence, any device having a GPS and a THM can use the \ourmethod{} web-API to provide real-time indicative AQI information as an add-on service.

\section{Methodology}
\label{Method}
This section describes the feature engineering followed by the AQI annotation module of \ourmethod{}. The details are as follows.

\subsection{Feature Extraction}
We aim to formulate an annotation model to get the AQI levels with user inputs through a THM/low-cost AQMDs/AQMS. As discussed in~\cite{zheng2015forecasting}, the air quality of a region depends on various meteorological parameters and geo-spatial features of the area.
\textcolor{\blue}{Moreover, we observe a significant dependency of air quality with time and consider including temporal features along with the aforementioned features that are extracted from different publicly available sources. The features are as follows:}

\begin{table}[!htbp]
\centering
\scriptsize
\caption{Activity-based Clustering to account for the Diurnal Variation}
\label{tab:time_cluster}
\begin{tabular}{|c|l|}
\hline
\multicolumn{1}{|c|}{\textbf{\begin{tabular}[c]{@{}c@{}}Cluster\\ Label\end{tabular}}} & \multicolumn{1}{c|}{\textbf{\begin{tabular}[c]{@{}c@{}}Local Time\\ +0530 UTC\end{tabular}}} \\ \hline
C\textsubscript{1}       & 00:00 - 06:59 \\ \hline
C\textsubscript{2}       & 07:00 - 09:59 \\ \hline
C\textsubscript{3}       & 10:00 - 16:59 \\ \hline
C\textsubscript{4}       & 17:00 - 23:59 \\ \hline
\end{tabular}
\end{table}

\textbf{(1) Meteorological Data:}
We primarily use the temperature and humidity measures obtained from AQMDs. Additionally, we crawl different meteorological parameters including feels like, pressure, wind speed, wind direction, rain, and cloud coverage from publicly available sources, Meteoblue\footnote{\url{www.meteoblue.com} (Accessed: \today)} and Open Weather \footnote{\url{https://openweathermap.org/} (Accessed: \today)}. While temperature and humidity show variations even within a city (\figurename{ \ref{dT_dH_city}}), the parameters that we crawl from the web typically remain similar over a larger region\footnote{There can be localized rainfalls, but they are typically for a concise duration and do not impact the AQI much.}. Therefore, the public sources give a good estimate of those parameters.

\begin{figure}[!htbp]
  \centering
  \includegraphics[width=0.90\linewidth]{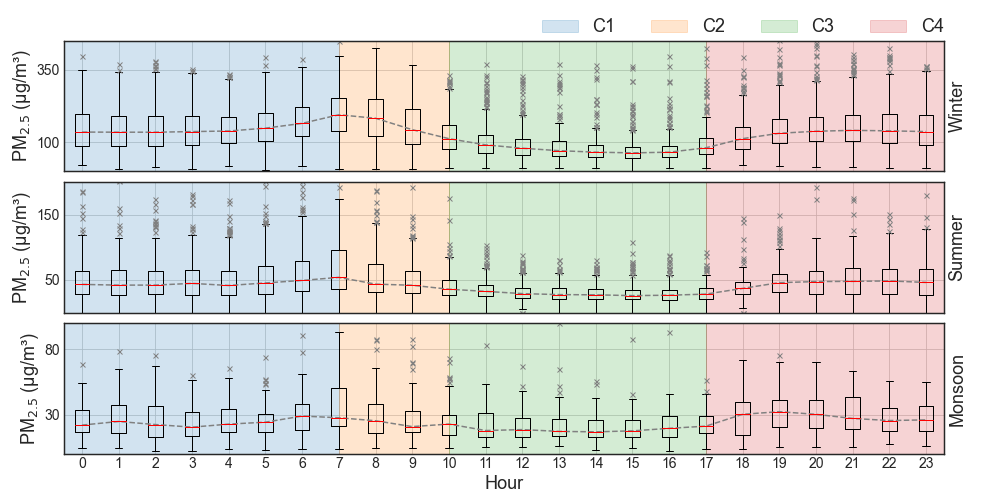}
  \caption{Relationship between Temporal features like Activity-based Clusters and Seasons (Summer, Monsoon, \& Winter) with PM\textsubscript{2.5}.}
  \label{temporal_proof}
\end{figure}


\textcolor{\blue}{\textbf{(2) Temporal Features:}
Air quality changes periodically, so it is critical to capture its temporal aspect. The temporal aspect we are most interested are \textit{diurnal variation}, \textit{seasonal variation} \& \textit{changes in urban transportation habits.} Hence, we introduce the features \textit{Hour of Day}, \textit{Activity-based Cluster}, \textit{Month}, \textit{Season}, \& \textit{Day of Week}. Hour of Day contributes majorly in diurnal variation as shown in \figurename{ \ref{temporal_proof}}.
As traffic activity influences the air quality at any location, based on our understanding of the activity level in the cities, we further split the day into multiple activity clusters, see \tablename~\ref{tab:time_cluster}. For example, very less activity can be observed from late at night till 07:00 in the morning. However, we see an increase in activity level at around 09:00 and 19:00 as people usually go to their workplaces and return home at these times. Again, a similar lack of activity can be observed during the noon and afternoon hours as people are in offices. Refer to \figurename{ \ref{temporal_proof}} to see a similar trend in real-world PM\textsubscript{2.5} data.} 

\textcolor{\blue}{We observe the relationship between different months of the year and air quality. Moreover, we also take three major seasons seen in India (i.e., winter, summer, and monsoon) and observe that the average pollution level is at its peak in winter, followed by summer and monsoon, as shown in \figurename{ \ref{temporal_proof}}. In winter, due to low humidity during the day and smog in the early morning, PM\textsubscript{2.5} rises. In summer, the humidity is relatively higher, and a significant amount of particles bind with the excess water in the air and precipitate on the ground. Finally, in the monsoon, when the humidity is at its peak, we observe the lowest amount of pollution.} 

\textcolor{\blue}{Including the above temporal features adds significant information to the set of existing input features and helps the model to learn complex temporal relationships for predicting the air quality at any location.}

\textbf{(3) Points of Interest (PoI):}
We use PoI extracted through Google's static-map API~\cite{10.1145/3549548}, which includes industrial locations, parks, urban areas, city centers, etc. These features give the spatial information that helps the model predict the AQI of a location. Static-maps API takes a geographical coordinate (i.e., latitude, longitude) and a zoom level to obtain the image of the spatially distributed PoIs around that coordinate. During the process, we poll the API with a predefined color coding scheme so that all the different PoIs appear in different colors. Finally, when the complete detailed image is obtained, we apply pixel-wise color filters to compute the percentage of the PoIs at each AQMD site in the studied area. 

\textbf{(4) Road Networks:}
The road network has a significant role in measuring air quality as vehicles are one of the primary contributors to pollution. We restrict the road types into three categories, viz. \textit{Highways}, \textit{Two-way roads}, and \textit{One-way roads}. The road type information is extracted from the Google Map API. The percentage of these different road types at each location is estimated using the same imaging techniques used for PoI detection.


\begin{figure}[!htbp]
    \centering
    \includegraphics[width=0.7\columnwidth]{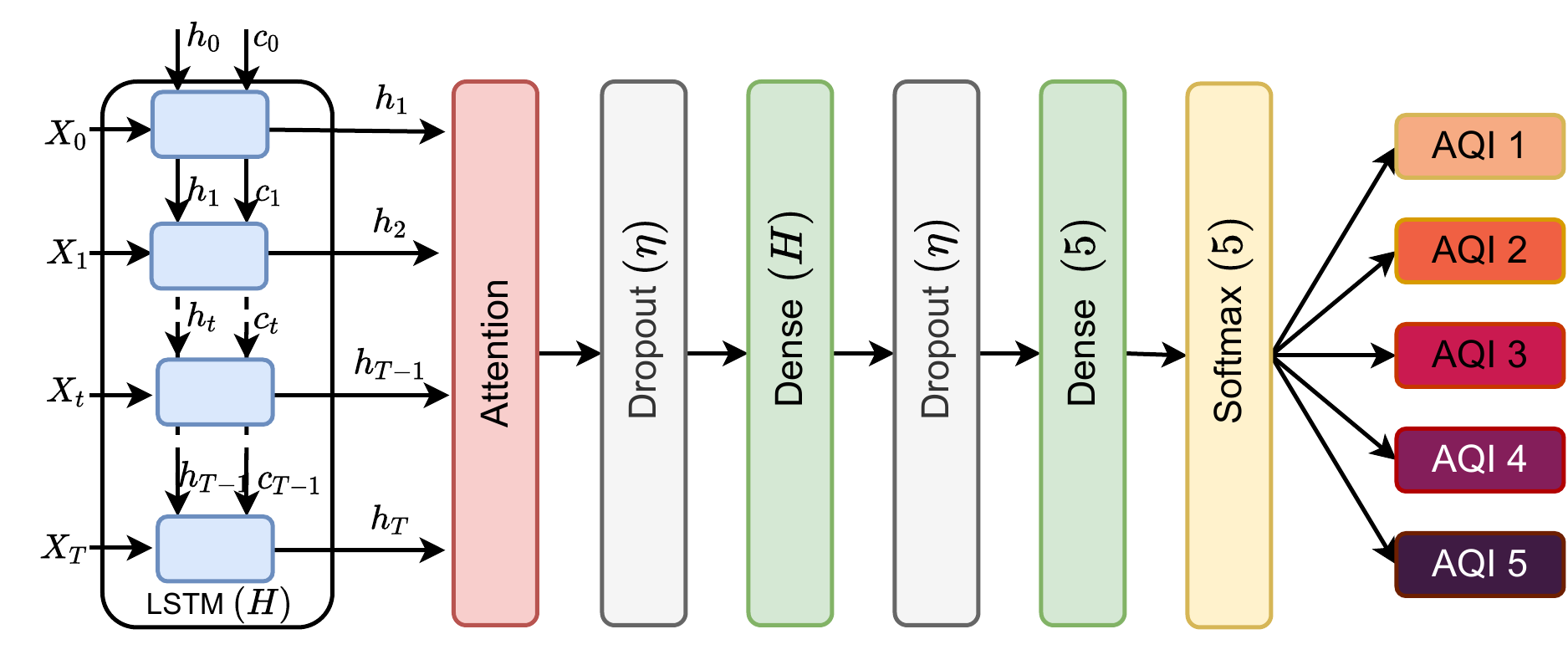}
    \caption{The LSTM-based revised model}
    \label{rnn_model}
\end{figure}

\subsection{Feature Selection and Pre-Training} 
In this work, we have collected $16$ features from the above four feature groups as follows. $\mathbf{f_1}$: \textit{Temperature},  $\mathbf{f_2}$: \textit{Humidity}, $\mathbf{f_3}$: \textit{Feels like}, $\mathbf{f_4}$: \textit{Hourly Minimum Temperature}, $\mathbf{f_5}$: \textit{Hourly Maximum Temperature}, $\mathbf{f_6}$: \textit{Pressure}, $\mathbf{f_7}$: \textit{Wind Speed}, $\mathbf{f_8}$: \textit{Wind Direction},  $\mathbf{f_9}$: \textit{Average Hourly Precipitation}, $\mathbf{f_{10}}$: \textit{Percentage Cloud Cover}, $\mathbf{f_{11}}$: \textit{Weather Type} like cloudy, sunny, hazy, overcast, thunderstorm etc., \textcolor{\blue}{$\mathbf{f_{12}}$: \textit{Hour of Day}, $\mathbf{f_{13}}$: \textit{Activity-based cluster}, $\mathbf{f_{14}}$: \textit{Month}, $\mathbf{f_{15}}$: \textit{Season} such as Winter, Summer, Monsoon, and $\mathbf{f_{16}}$: \textit{Day of Week}. The spatial similarity is measured using the ten spatial parameters, namely road types like one-way, two-way, and highway, PoI types like human-made structures, natural land, educational institutes, medical institutes, water bodies, parks, shopping malls, and other attractions.} 


\textcolor{\blue}{As we explained earlier, we create separate models for each city as they may possess different environmental and climate characteristics. Air quality at any location depends on the past few hours of meteorology and weather. Therefore, besides the newly added temporal features ($\mathbf{f_{12}}$ to $\mathbf{f_{16}}$), we also incorporated data for the past $T$ hours, which is referred to as window size in the rest of the paper. The value of window size $T$ is experimentally obtained in Section~\ref{Eval}. Subsequently, the collective series of input features are used to annotate the air quality index of the $T^{th}$ timestamp.}


\textcolor{\blue}{Realizing the opportunity for parameter sharing across timestamps, we utilize a Long Short Term Memory (LSTM)~\cite{yu2019review} layer, with $H$ neurons, that helps to learn the spatio-temporal relationship between the features with a minimal set of trainable weights. For our model, the LSTM layer works as a sequence-to-sequence non-linear transform, which is further reduced to a crisped vector with the help of the Bahdanau Attention layer~\cite{chorowski2015attention}. The Attention layer is used to filter unnecessary information from the temporal data and focus on specific temporal events to compress the entire sequence to a denser vector-representation. The output of the Attention layer is finally fed to the neural network classifier. The classifier consists of two hidden layers with $H$ neurons and dropout with rate $\eta$ before each of the hidden layers to reduce overfitting. Finally, the output layer of the classifier uses softmax activation to yield a joint probability distribution over 5 AQI classes. The above LSTM-based AQI annotation model is shown in \figurename{ \ref{rnn_model}}. The model is trained with the set of hyper-parameters listed in Section~\ref{Eval} to minimize the categorical cross-entropy loss.}


\subsection{Real-time Annotation}
\textcolor{\blue}{We propose a framework that yields a specialized annotation of air quality in a region given a generic and readily available set of features.} Such a system can be realized in the following way. A user who has deployed a low-cost device with partial sensing capabilities that measure temperature and humidity can utilize \ourmethod{} to get the AQI annotation at that location. However, we assume that the device is calibrated correctly and has an acceptable degree of precision and accuracy. The system also collects the GPS coordinates of the deployed device to crawl the remaining modalities from the publicly available weather APIs and compute the spatial distribution of the surroundings, which are then fed to the pre-trained model of \ourmethod{} to get the data annotated. 

\section{Evaluation}
\label{Eval}
For a detailed evaluation with the available ground truth, we choose a set of AQMDs for evaluating the performance of \ourmethod{} by considering disjoint held-out sets of devices. We did not use the full sensing capabilities for these held-out devices and only considered temperature and humidity as the input sensing modalities. Additionally, we extract the remaining features from the GPS coordinates of the corresponding devices and the timestamp of the sensor values. Concerning the AQI ground-truth, we compute the index from the $PM_{2.5}$ data available from the specialized sensors attached to these devices. Subsequently, the labeled AQI is compared with the ground truth to evaluate the framework's performance.

\subsection{Implementation and Experimental Setup}

This section describes the implementation details and the experimental setup to make the model reproducible. The section is organized as follows. We mention the implementation details in the following subsection and move on to the experimental details, where we revisit the data distribution, clarify our evaluation metric, and quantify the model training time in the particular setup.

\begin{table}[!htbp]
\centering
\scriptsize
\caption{The Hyper-parameters details of the LSTM-based model used for the final pre-trained model}
\label{hparams}
\begin{tabular}{|l|c|l|c|} 
\hline
\textbf{Parameter}    & \textbf{Value} & \textbf{Parameter}    & \textbf{Value}  \\ 
\hline
Hidden units ($H$)    & 128            & Optimizer ($\nabla$)             & Adam            \\ 
\hline
Dropout rate ($\eta$) & 0.2            & Learning rate ($\alpha$) & 0.001           \\ 
\hline
Activation ($\sigma$) & tanh           & Epoch ($E$)             & 1000            \\ 
\hline
l2 loss coef ($\lambda$)   & 0.001          & Batch size ($B$)        & 256             \\
\hline
\end{tabular}
\end{table}

	\subsubsection{Implementation Details and Baselines:}

\textcolor{\blue}{As a baseline, we train a Random Forest model~\cite{breiman2001random} (referred to as RF) to create the city-specific models. Moreover, we train the Random Forest model with the newly added temporal features (referred to as RF+T) to understand the direct contribution of such features in improving the quality of annotation.}
\textcolor{\blue}{To implement the Random Forest baselines, we use the number of estimators to be 100 and the maximum depth of each tree to be 20. The best set of hyper-parameters for the LSTM-based model is obtained with grid search in our dataset and is mentioned in \tablename~\ref{hparams}.}



\subsubsection{Experimental Setup:}
\label{ExpSet}
\textcolor{\blue}{Here we explain the experimental setup and metrics for the evalution.}


\begin{table}[!htbp]
\centering
\scriptsize
\caption{AQI data availability for \CityA{} and \CityB{} and it's class-wise distribution}
\label{data_details}
\begin{tabular}{|l|l|l|l|l|l|l|l|l|} 
\hline
\textbf{City} & \textbf{Duration} & \textbf{Devices} & \textbf{AQI 1} & \textbf{AQI 2} & \textbf{AQI 3} & \textbf{AQI 4} & \textbf{AQI 5} & \textbf{Total Data}  \\ 
\hline
\CityA{}        & 12 months               & 4                & 3405              & 3810              & 2538              & 1570              & 4222              & 15545                    \\ 
\hline
\CityB{}        & 17 months                & 12               & 29515              & 41427              & 22495              & 13221              & 39385              & 146043                    \\
\hline
\end{tabular}
\end{table}

\textcolor{\blue}{\textbf{Data Distribution}: As we can see from \tablename~\ref{data_details}, for \CityA{}, despite the data availability of 12 months, having 4 AQMDs yields around 15.5K data points compared to \CityB{}, which has data for 17 months from 12 AQMS, yields 146K data points. This city-wise data distribution actually brings out the data-volume based dependencies of models, as shown in the following sections.}

\textcolor{\blue}{\textbf{Evaluation Metric}: To evaluate the models, we have considered the F1-score (i.e., F-measure) to be the performance measure of the model as it is considered the standard for classification model evaluation. In our case, we consider the weighted-averaged F1-score. The weighted F1-score is calculated by taking the mean of all per-class F1-scores while considering each class's support, where support refers to the number of actual occurrences of the class in the training dataset.}

\textcolor{\blue}{\textbf{Setup}: The RF, RF + Temporal \& LSTM-based models were trained on a standard laptop (with 8GB primary memory running MacOS v12.0.1 with base-kernel version: 21.1.0) and for the hyper-parameter tuning we utilized Colab (2 x vCPU, GPU Nvidia Tesla K80, \& RAM 12GB). The training time for the individual models is given here RF: 26 seconds, RF + Temporal: 26.23 seconds, \& LSTM-based: 5 hours. We have used the software package based on \texttt{Python3.8.12}, \texttt{Tensorflow v2.4.1}, and  \texttt{Scikit-learn v1.0.2} for the implementation.}

\subsection{ Impact of Window Size}

\begin{figure}[!htbp]
     \centering
        \includegraphics[width=0.5\linewidth,keepaspectratio]{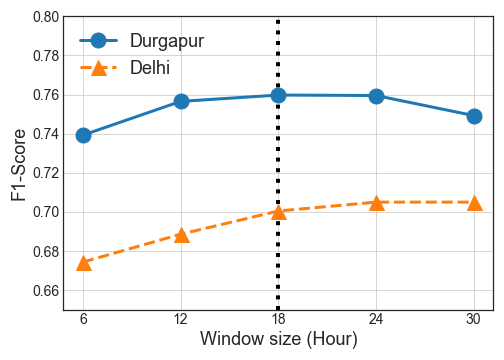}
        \caption{Evaluation \& Selection of Optimal Window Size (\textit{T})}
        \label{window_size}
\end{figure}

\textcolor{\blue}{In this temporal model, we use a windowed input pattern to provide the necessary time steps for the input feature. In this section, we analyze the impact of window size on the final model performance. As shown in \figurename{ \ref{window_size}}, both \CityA{} and \CityB{} achieves a consistent F1-score with a window size of $18$ hours. Based on this observation, we consider this window size = $18$ hours for the remaining experiments. }
		
\begin{table}[!htbp]
\scriptsize
\centering
\caption{Leave-one-out, spatial similarity, and distance based AQI annotation F1-score in \CityA{} with Random Forest (RF), Random Forest with Temporal features (RF+T), and LSTM-based Model (LSTM)}
\label{poi_dis_city1}
\begin{tabular}{|l|l|l|l|l|l|l|l|l|l|l|l|} 
\hline
\multicolumn{1}{|c|}{\multirow{2}{*}{\begin{tabular}[c]{@{}c@{}}\textbf{ Test }\\\textbf{Device }\end{tabular}}} & \multicolumn{3}{c|}{\textbf{Leave-one-out}}                                                                                                                                  & \multicolumn{4}{c|}{\textbf{Spatial Similarity }}                                                          & \multicolumn{4}{c|}{\textbf{Distance}}                                                                      \\ 
\cline{2-12}
\multicolumn{1}{|c|}{}                                                                                           & \multicolumn{1}{c|}{\textcolor[rgb]{0.2,0.2,0.2}{\textbf{RF}}} & \multicolumn{1}{c|}{\textcolor[rgb]{0.2,0.2,0.2}{\textbf{RF+T}}} & \multicolumn{1}{c|}{\textbf{LSTM}}  & \multicolumn{1}{c|}{\textbf{Train~AQMD}} & \multicolumn{1}{c|}{\textbf{RF}} & \textbf{RF+T} & \textbf{LSTM} & \multicolumn{1}{c|}{\textbf{Train~AQMD}} & \multicolumn{1}{c|}{\textbf{RF}} & \textbf{RF+T} & \textbf{LSTM}  \\ 
\hline
AQMD 1                                                                                                           & \textcolor[rgb]{0.2,0.2,0.2}{0.70}                             & \textcolor[rgb]{0.2,0.2,0.2}{\textbf{0.74}}                      & \textcolor[rgb]{0.2,0.2,0.2}{0.72} & AQMD 3                                   & 0.63                             & 0.66          & 0.63         & AQMD 3                                   & 0.63                             & 0.66          & 0.63          \\ 
\hline
AQMD 2                                                                                                           & \textcolor[rgb]{0.2,0.2,0.2}{0.68}                             & \textcolor[rgb]{0.2,0.2,0.2}{\textbf{0.71}}                      & \textcolor[rgb]{0.2,0.2,0.2}{0.67} & AQMD 3                                   & 0.63                             & \textbf{0.66} & 0.62         & AQMD 4                                   & 0.58                             & 0.61          & 0.57          \\ 
\hline
AQMD 3                                                                                                           & \textcolor[rgb]{0.2,0.2,0.2}{0.74}                             & \textcolor[rgb]{0.2,0.2,0.2}{\textbf{0.76}}                      & \textcolor[rgb]{0.2,0.2,0.2}{0.75} & AQMD 2                                   & 0.67                             & \textbf{0.73} & 0.71         & AQMD 4                                   & 0.66                             & 0.70          & 0.66          \\ 
\hline
AQMD 4                                                                                                           & \textcolor[rgb]{0.2,0.2,0.2}{0.69}                             & \textcolor[rgb]{0.2,0.2,0.2}{\textbf{0.70}}                      & \textcolor[rgb]{0.2,0.2,0.2}{0.69} & AQMD 3                                   & 0.63                             & 0.63          & 0.62         & AQMD 3                                   & 0.63                             & 0.63          & 0.62          \\
\hline
\end{tabular}
\end{table}

\begin{table}[!htbp]
\scriptsize
\centering
\caption{Leave-one-out, spatial similarity, and distance based AQI annotation F1-score in \CityB{} with Random Forest (RF), Random Forest with Temporal features (RF+T), \& LSTM-based Model (LSTM)}
\label{poi_dis_city2}
\begin{tabular}{|l|l|l|l|l|l|l|l|l|l|l|l|} 
\hline
\multicolumn{1}{|c|}{\multirow{2}{*}{\begin{tabular}[c]{@{}c@{}}\textbf{Test }\\\textbf{Device}\end{tabular}}} & \multicolumn{3}{c|}{\textbf{Leave-one-out}}                                                                                                                                           & \multicolumn{4}{c|}{\textbf{Spatial Similarity}}                                                            & \multicolumn{4}{c|}{\textbf{Distance}}                                                                       \\ 
\cline{2-12}
\multicolumn{1}{|c|}{}                                                                                         & \multicolumn{1}{c|}{\textcolor[rgb]{0.2,0.2,0.2}{\textbf{RF}}} & \multicolumn{1}{c|}{\textcolor[rgb]{0.2,0.2,0.2}{\textbf{RF+T}}} & \multicolumn{1}{c|}{\textbf{LSTM}}           & \multicolumn{1}{c|}{\textbf{Train~AQMS}} & \multicolumn{1}{c|}{\textbf{RF}} & \textbf{RF+T} & \textbf{LSTM}  & \multicolumn{1}{c|}{\textbf{Train~AQMS}} & \multicolumn{1}{c|}{\textbf{RF}} & \textbf{RF+T} & \textbf{LSTM}   \\ 
\hline
AQMS 1                                                                                                         & \textcolor[rgb]{0.2,0.2,0.2}{0.58}                             & \textcolor[rgb]{0.2,0.2,0.2}{0.59}                               & \textcolor[rgb]{0.2,0.2,0.2}{\textbf{0.63}} & AQMS 8                                   & 0.52                             & 0.55          & 0.55          & AQMS 8                                   & 0.52                             & 0.55          & 0.55           \\ 
\hline
AQMS 2                                                                                                         & \textcolor[rgb]{0.2,0.2,0.2}{0.69}                             & \textcolor[rgb]{0.2,0.2,0.2}{0.69}                               & \textcolor[rgb]{0.2,0.2,0.2}{\textbf{0.70}} & AQMS 10                                  & 0.46                             & 0.47          & 0.49          & AQMS 4                                   & 0.62                             & \textbf{0.68} & 0.67           \\ 
\hline
AQMS 3                                                                                                         & \textcolor[rgb]{0.2,0.2,0.2}{0.58}                             & \textcolor[rgb]{0.2,0.2,0.2}{0.58}                               & \textcolor[rgb]{0.2,0.2,0.2}{\textbf{0.61}} & AQMS 8                                   & 0.48                             & 0.52          & 0.50          & AQMS 6                                   & 0.55                             & \textbf{0.59} & 0.58           \\ 
\hline
AQMS 4                                                                                                         & \textcolor[rgb]{0.2,0.2,0.2}{\textbf{0.70}}                    & \textcolor[rgb]{0.2,0.2,0.2}{0.69}                               & \textcolor[rgb]{0.2,0.2,0.2}{0.68}          & AQMS 3                                   & 0.46                             & 0.50          & 0.45          & AQMS 5                                   & 0.63                             & \textbf{0.73} & 0.68           \\ 
\hline
AQMS 5                                                                                                         & \textcolor[rgb]{0.2,0.2,0.2}{0.71}                             & \textcolor[rgb]{0.2,0.2,0.2}{0.70}                               & \textcolor[rgb]{0.2,0.2,0.2}{\textbf{0.73}} & AQMS 9                                   & 0.59                             & 0.62          & 0.64          & AQMS 4                                   & 0.61                             & \textbf{0.67} & \textbf{0.67}  \\ 
\hline
AQMS 6                                                                                                         & \textcolor[rgb]{0.2,0.2,0.2}{0.61}                             & \textcolor[rgb]{0.2,0.2,0.2}{0.61}                               & \textcolor[rgb]{0.2,0.2,0.2}{\textbf{0.63}} & AQMS 9                                   & 0.57                             & \textbf{0.60} & \textbf{0.60} & AQMS 3                                   & 0.52                             & 0.57          & 0.55           \\ 
\hline
AQMS 7                                                                                                         & \textcolor[rgb]{0.2,0.2,0.2}{0.71}                             & \textcolor[rgb]{0.2,0.2,0.2}{0.71}                               & \textcolor[rgb]{0.2,0.2,0.2}{\textbf{0.73}} & AQMS 1                                   & 0.52                             & 0.56          & 0.56          & AQMS 8                                   & 0.58                             & \textbf{0.63} & 0.62           \\ 
\hline
AQMS 8                                                                                                         & \textcolor[rgb]{0.2,0.2,0.2}{0.67}                             & \textcolor[rgb]{0.2,0.2,0.2}{0.67}                               & \textcolor[rgb]{0.2,0.2,0.2}{\textbf{0.71}} & AQMS 1                                   & 0.50                             & 0.52          & 0.53          & AQMS 1                                   & 0.50                             & 0.52          & 0.53           \\ 
\hline
AQMS 9                                                                                                         & \textcolor[rgb]{0.2,0.2,0.2}{0.69}                             & \textcolor[rgb]{0.2,0.2,0.2}{0.70}                               & \textcolor[rgb]{0.2,0.2,0.2}{\textbf{0.72}} & AQMS 8                                   & 0.58                             & 0.64          & 0.62          & AQMS 8                                   & 0.58                             & 0.64          & 0.62           \\ 
\hline
AQMS 10                                                                                                        & \textcolor[rgb]{0.2,0.2,0.2}{0.57}                             & \textcolor[rgb]{0.2,0.2,0.2}{0.58}                               & \textcolor[rgb]{0.2,0.2,0.2}{\textbf{0.61}} & AQMS 11                                  & 0.59                             & 0.62          & 0.62          & AQMS 11                                  & 0.59                             & 0.62          & 0.62           \\ 
\hline
AQMS 11                                                                                                        & \textcolor[rgb]{0.2,0.2,0.2}{0.65}                             & \textcolor[rgb]{0.2,0.2,0.2}{0.66}                               & \textcolor[rgb]{0.2,0.2,0.2}{\textbf{0.68}} & AQMS 10                                  & 0.62                             & 0.63          & 0.63          & AQMS 10                                  & 0.62                             & 0.63          & 0.63           \\ 
\hline
AQMS 12                                                                                                        & \textcolor[rgb]{0.2,0.2,0.2}{0.69}                             & \textcolor[rgb]{0.2,0.2,0.2}{0.68}                               & \textcolor[rgb]{0.2,0.2,0.2}{\textbf{0.72}} & AQMS 1                                   & 0.50                             & 0.52          & 0.55          & AQMS 5                                   & 0.62                             & \textbf{0.68} & \textbf{0.68}  \\
\hline
\end{tabular}
\end{table}

\subsection{AQI Annotation Performance}

The performance of \ourmethod{} for AQI annotation at \CityA{}, and \CityB{} is shown in \tablename~\ref{poi_dis_city1}, and \tablename~\ref{poi_dis_city2} respectively. We evaluate \ourmethod{} in three types of settings: (i) leave-one-out for the understanding of the real-world performance of the framework at unseen locations, (ii) Spatial similarity, and (iii) Distance-based similarity wise personalized to observe the impact of geolocation-based proximity on the choice of AQMD (or AQMS). The primary observations are as follows.

In general, we see a significant improvement in AQI annotation quality over the base Random Forest model after adding the temporal features in both \CityA{} and \CityB{}. Therefore, it is evident that the temporal features add value to the proposed framework. However, we can see in \tablename~\ref{poi_dis_city1} that the Random Forest model with temporal features outperforms the LSTM-based model for \CityA{}. The primary reason is the limited amount of available data at \CityA{} (see \tablename~\ref{data_details}). Notably, for \CityB{}, where we have a relatively larger dataset, a significant improvement can be observed using the LSTM-based model. The improvement can be observed in \tablename~\ref{poi_dis_city2} with leave-one-out based testing.

\begin{figure}[!htbp]
        \captionsetup[subfigure]{}
        \begin{center}
            \subfloat[\label{roc_good1}]{
                    \includegraphics[width=0.45\linewidth,keepaspectratio]{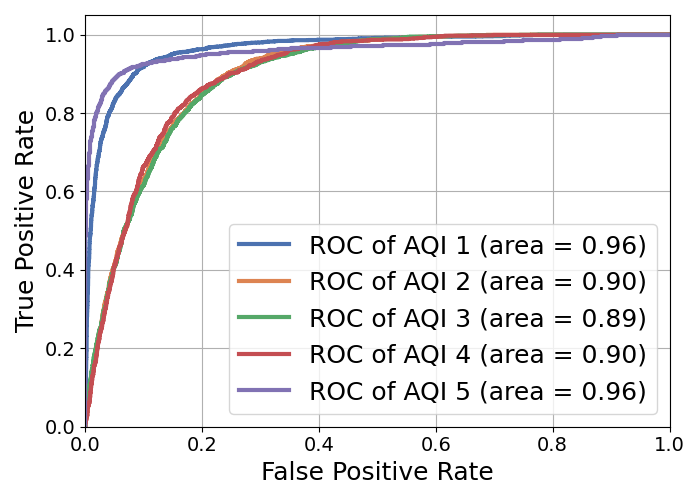}
            }
            \subfloat[\label{roc_good2}]{
                    \includegraphics[width=0.45\linewidth,keepaspectratio]{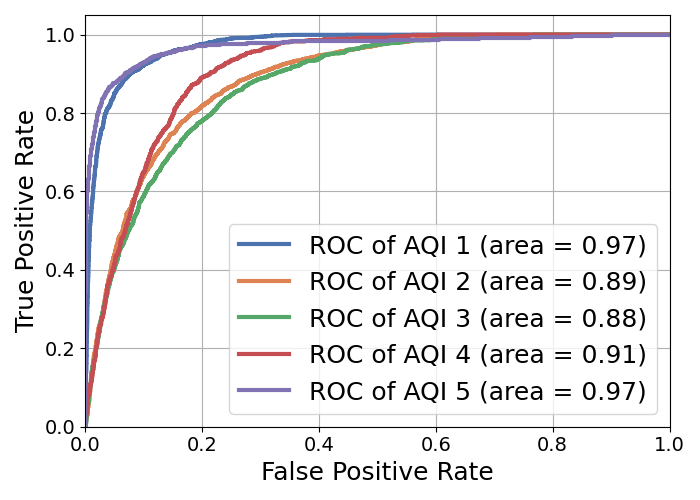}
            }\
            \subfloat[\label{roc_bad1}]{
                    \includegraphics[width=0.45\linewidth,keepaspectratio]{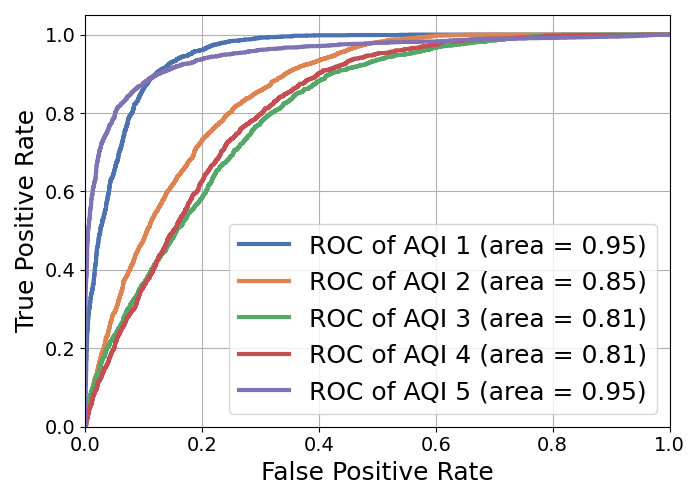}
            }
            \subfloat[\label{roc_bad2}]{
                    \includegraphics[width=0.45\linewidth,keepaspectratio]{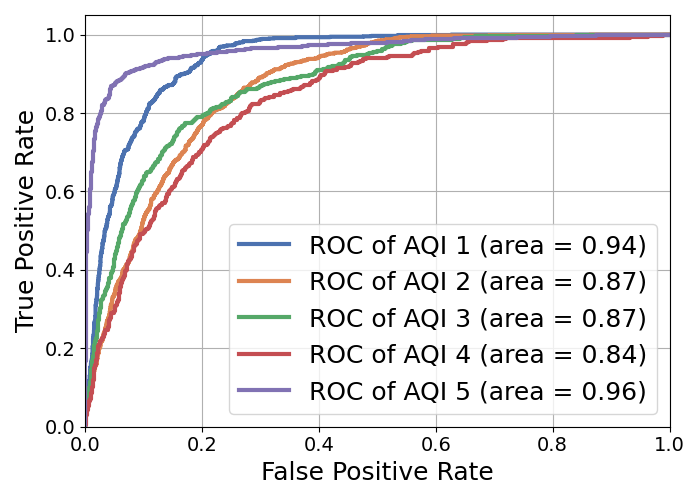}
            }
        \end{center}
        \caption{Annotation AUC-ROC in - \CityB{} (a) AQMS 8 (b) AQMS 12 (c) AQMS 3, \CityA{} (d) AQMD 2}
        \label{roc_plots}
\end{figure}

For the leave-one-out experiments in \CityA{}, the three models are trained on the data from 3 AQMDs, while the remaining AQMD is used as a testing device, see \tablename~\ref{poi_dis_city1}. Similarly, we evaluate the performance for AQI annotation at \CityB{}, using 11 out of 12 state-deployed AQMSs to train the annotation model and test the framework on the remaining one, as shown in \tablename~\ref{poi_dis_city2}. Moreover, we show the performance of the LSTM-based model in contrast with the baseline Random Forest (RF) and Random Forest with temporal features (RF+T) model. As mentioned earlier, in \CityB{}, we observe that the LSTM-based model shows appreciable improvement in performance over the baseline models. While in \CityA{}, the Random Forest with temporal features scales well as we do not have enough data for \CityA{} to take advantage of deep learning. 
\figurename{ \ref{roc_plots}} shows the individual ROC plots for all AQI classes for some of the devices in the leave-one-out setting. We can observe that the AUC of AQI class 5 across \figurename~\ref{roc_plots} is the highest, followed by AQI class 1, and then AQI class 2 to AQI class 4, which indicates more misclassifications for these AQI classes. We further analyze the severity of this misclassification in the next section. 

In spatial and distance-based similarity experiments, we train the annotation model with the most spatially similar or based on proximity in terms of geolocation AQMD (or AQMS). As per the results shown in \tablename~\ref{poi_dis_city1} and \tablename~\ref{poi_dis_city2} for these similarity-based experiments, we observe that there is little scope for improvement for similarity-based pre-training due to the inherent data scarcity problem and city-specific heterogeneity. We further analyze these experiments in the following sections.

Summarily, from the above experiments, we can clearly observe that the newly added temporal features improve the annotation quality in both \CityA{} and \CityB{}, while the LSTM-based model shows superior performance as compared to the baseline models, provided that we have enough data for a particular city.

\begin{figure}[!htbp]
        \captionsetup[subfigure]{}
        \begin{center}
            \subfloat[\label{aqi_DGP}]{
                    \includegraphics[width=0.45\linewidth,keepaspectratio]{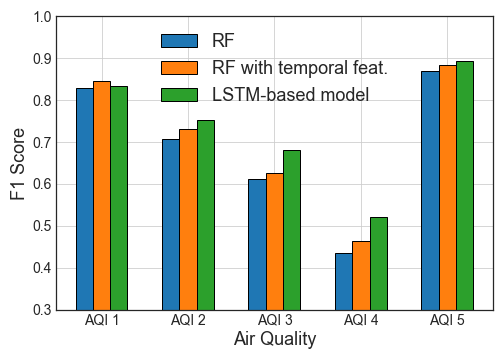}
            }
            \subfloat[\label{aqi_DELHI}]{
                    \includegraphics[width=0.45\linewidth,keepaspectratio]{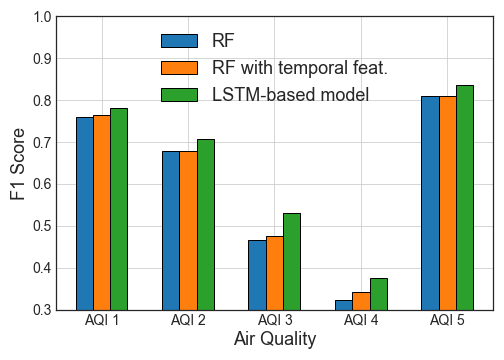}
            }
        \end{center}
        \caption{AQI Annotation Performance using RF, RF + Temporal \& LSTM-based models across -(a) \CityA{} \& (b) \CityB{} using F1-score}
        \label{aqi_wise_perf}
\end{figure}

\subsection{Framework Insights}
In this section, we dissect the framework to further analyze the results and several associated factors. The analysis is described as follows.
\subsubsection{Class-wise Performance:}
\textcolor{\blue}{We next analyze the AQI-wise annotation performance of the LSTM-based model and the baselines for both cities. We compute the F1-score for annotating individual AQIs for the held-out devices over all the four combinations in \CityA{} (3 for train and 1 for the test) and 12 combinations in \CityB{} (11 for train and 1 for the test), and then averaged over all the devices across AQI levels. \figurename{ \ref{aqi_wise_perf}} shows the performance in annotating individual AQIs. Interestingly we observe that across both the cities, AQI 3 and 4 consistently suffer the most. Indeed, the low F1-score in correctly predicting AQI 3 and 4 affects the overall performance of \ourmethod{}, as we have seen earlier in \tablename~\ref{poi_dis_city1} and \tablename~\ref{poi_dis_city2}. By revisiting \figurename{ \ref{th_aqi_sp}}, we observe that the individual clusters for AQIs 1, 2, and 5 have low overlap among themselves; however, AQIs 3 and 4 have a more extensive spread and overlap with clusters of other AQIs, which confuses the model.} 

\subsubsection{Model Severity Analysis:}

\begin{figure}[!htbp]
    \captionsetup[subfigure]{}
    \begin{center}
        \subfloat[\label{class_4_aqi_city1}]{
                \includegraphics[width=0.45\linewidth,keepaspectratio]{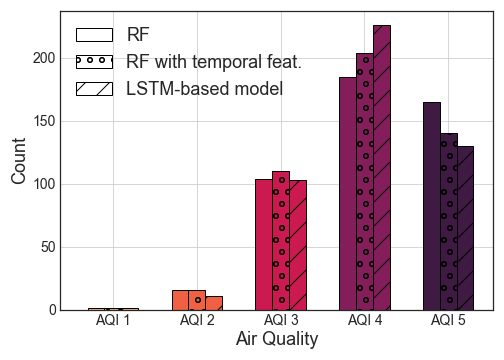}
        }
        \subfloat[\label{class_4_aqi_city2}]{
                \includegraphics[width=0.45\linewidth,keepaspectratio]{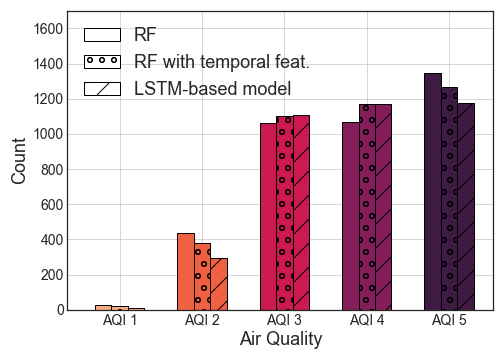}
        }
    \end{center}
    \caption{Performance analysis to measure the classification of target AQI Class 4  -(a) \CityA{} \& (b) \CityB{} to understand the misclassification distribution across other AQI classes with respect to RF, RF + Temporal \& LSTM-based models.}
    \label{Class_4_preds}
\end{figure}
\textcolor{\blue}{Here, we present the erroneous prediction cases for AQI 4 as a case study as AQI 4 consistently suffers the most across both the cities, as seen in \figurename{ \ref{annotation_quality}}. \tablename~\ref{data_details} shows that AQI 4 is available the least in both \CityA{} \& \CityB{}.  So it becomes important to understand the performance of the models when used in class 4 to figure out where it is misclassified and if the LSTM-based model can improve upon the baselines. \figurename{ \ref{Class_4_preds}} shows the distribution of predicted AQI levels when the true AQI level is 4. From the figure, we can observe that all models wrongly annotate AQI class $4$ as AQI class $3$ or $5$, with most cases getting labeled as AQI $5$. Although this accounts for a misclassification, the model actually predicts an AQI class that has a higher severity in terms of air quality. For example, in this case, the AQI class $4$ is mostly predicted as class $5$, which is more severe and hardly mislabels it as AQI $1$ or $2$. Thus, the quality of annotations generated by the model maintains a strict standard without compromising the severity of the AQI classes.} 

\subsubsection{Spatial Analysis:}

\textcolor{\blue}{From the preliminary observations (See Section~\ref{dataset}), we understand that the overall AQI distribution changes with the location change. This, in turn, poses a bigger question -- whether there should be a generalized global pre-trained model for each city, and, if not, how to choose a pre-trained model if there are multiple data sources (AQMDs or AQMSs) available in a city for the creation of more than one pre-trained model. Considering this, we analyze the performance of \ourmethod{} by designing multiple pre-trained models based on two different factors -- (a) similarity in the distribution of the spatial cluster and (b) physical distance between any pair of AQMDs or AQMSs. We analyze the results for \CityA{} in \tablename~\ref{poi_dis_city1} \& for \CityB{} in \tablename~\ref{poi_dis_city2} as follows. Let $\mathcal{A}_T$ be the test AQMD/AQMS. For case (a), we develop the pre-trained model using the data from the AQMD/AQMS that shows maximum spatial similarity with $\mathcal{A}_T$. Similarly, for case (b), the pre-trained model has been developed using the data from the AQMD/AQMS, which is physically closest to $\mathcal{A}_T$. \tablename~\ref{poi_dis_city1} and \tablename~\ref{poi_dis_city2} present the performance of \ourmethod{} considering these two factors for \CityA{} and \CityB{}, respectively. From the results, it is evident that for \CityA{}, \ourmethod{} gives better results if the spatial distribution similarity is considered while choosing the pre-trained model. However, for \CityB{}, this particular assumption does not hold, with the distance factor having a potentially stronger impact on the choice of the pre-trained model. The reasons behind such a change can be attributed to the diverse nature of the two cities. \CityA{} has several pollution-prone industrial areas, and thus, the places with similar spatial distribution provide a better estimate of the air quality.
On the contrary, \CityB{} has a larger area with similar spatial clutter, and therefore, places that are physically close to each other have similar environmental patterns. However, one critical observation that we gain is that none of these models performs better than the global leave-one-out model. Therefore, we argue that the global model works best for any city, although we need city-wise pre-training. }

\subsection{How Many Devices are Good Enough?}

\begin{figure}[!htbp]
    \centering
    \includegraphics[width=0.5\linewidth,keepaspectratio]{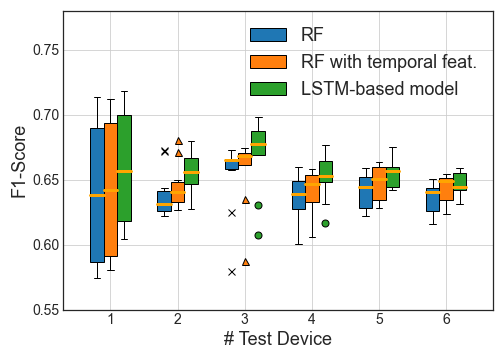}
    \caption{Performance analysis of RF, RF + Temporal \& LSTM-based models when decreasing the number of AQMS in \CityB{}. The x-axis shows the number of AQMS decreased from the pre-training}
    \label{box_plot_1by1}
\end{figure}
\textcolor{\blue}{Undoubtedly, the robustness of the final annotation model is highly dependent on the number of devices that provide the initial data to pre-train the model. In this section, we analyze the impact of sparsity in the device (AQMD or AQMS) deployment on the annotation quality of the pre-trained model. A shown in \figurename{ \ref{box_plot_1by1}}, the F1-score drops with decrease in total number of devices (here AQMSs deployed in \CityB). However, it is comforting for us to observe that the LSTM-based model still performs well in comparison to the other baselines with a consistent F1-score (weighted) of $\approx 65\%$ even with half the original deployment span (reduced to $6$ from $12$ AQMSs originally deployed in \CityB).}

\begin{figure}[!htbp]
        \captionsetup[subfigure]{}
        \begin{center}
            \subfloat[\label{dgp_online}]{
                    \includegraphics[width=0.45\linewidth,keepaspectratio]{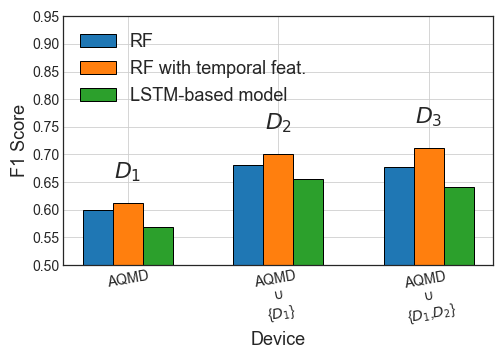}
            }
            \subfloat[\label{delhi_online}]{
                    \includegraphics[width=0.45\linewidth,keepaspectratio]{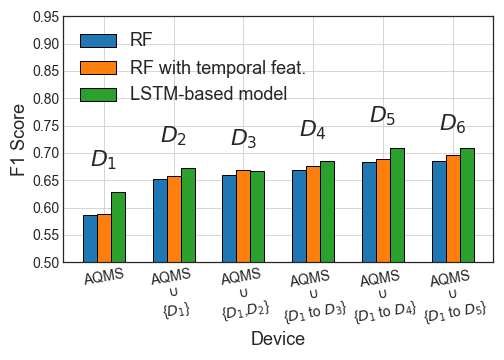}
            }
        \end{center}
        \caption{Impact on the performance of the pre-trained model with newly annotated data when progressively increasing the number of devices in  (a) \CityA{} \& (b) \CityB{}}
        \label{annotation_quality}
\end{figure}

\subsection{Impact of Progressive Device Deployment}

\textcolor{\blue}{The proposed method can be used to get the AQI of a location on a personalized scale using a low-cost THM and the GPS location. Following the usual sparse deployment of the AQMDs or AQMSs throughout any city, we further analyzed whether the annotated AQI data can augment the pre-trained model to enhance the quality of the annotation over time. This provides insight regarding improvement in F1-score over time. As depicted in \figurename{ \ref{annotation_quality}} where $D_i$ is the $i^{th}$ newly annotated device. Indeed, it is observed that in the case of \CityB{}, the annotation quality of the model improves with time and the addition of datasets from newly included devices. Similarly, for \CityA{} as well, we observe that the Random Forest with temporal features performs well, albeit here as well, the LSTM-based model performs poorly due to the lack of adequate data instances.}

\section{Limitations of the Current Framework \& Future Work}
\label{discussion}

Although \ourmethod{} can annotate the data with significant accuracy, the broader and more realistic applicability of such a framework possibly needs further investigation of some crucial factors. This section discusses the limitations of the current work and some of the future directions we plan to include in the subsequent versions of \ourmethod{}. The details are as follows.

\begin{itemize}
    \item[1. ] \textbf{Minimum devices required:} Each city's air quality behaves somewhat differently from others. The meteorology also varies from city to city. Thus, one of the primary requirements of \ourmethod{} for annotating the data from any new thermo-hygrometer at any given location is the fact that it needs a pre-trained model which has been created with the data available from one or more AQMSs (or AQMDs) available in that city (or area). Hence a target city where we want to deploy \ourmethod{} must have existing AQMSs or AQMDs to generate the initial data for model building. Undoubtedly, this restricts the applicability of \ourmethod{} as the availability of such infrastructures can be a major bottleneck. In Section~\ref{Eval}, we estimate the minimum number of devices required to get comparable annotation quality in \CityB{}. However, further studies must be done to have a holistic view of all cities.
    
    \item[2. ] \textbf{Optimal device placement:} In an underdeveloped area where no pre-deployment has happened, we need to deploy the AQMDs in such locations that efficiently capture the true distribution of the pollutants. The optimal placement of AQMDs across the cities is a major factor and can significantly impact the framework's performance. It generally includes the assessment of \textit{(a)} given some devices, how to best place them, and \textit{(b)} how many devices we need to place so that we can optimally estimate the AQI of the place in question. Air quality variability can be due to various factors like the geographic nature~\cite{gulia2020sensor, HOEK20087561}, variation in traffic patterns~\cite{matte2013monitoring, zheng2013u}, etc. Therefore, optimal site selection for sensor placement is a non-trivial problem that needs to be investigated in a detailed manner which we intend to perform in future versions.
    
    \item[3. ] \textbf{Few-Shot Label Annotation:} Section~\ref{Eval} shows that the annotation performance of \ourmethod{} suffers when less data is available in a city for building the initial model. This is particularly an issue because the current version of \ourmethod{} can not aggregate publicly available data from various cities to learn lower-level features. Instead, the framework treats each city separately. Thus, one of the major improvements in this direction would be to make the framework adapted to a city with limited or no prior knowledge of that city's environmental patterns.
\end{itemize}

\section{Conclusion}
\label{Conclude}
In this paper, we propose a framework named \ourmethod{} that can be used to annotate any dataset containing temperature and humidity generated through low-cost THM with AQI labels. In the backend, \ourmethod{} relies on a pre-trained LSTM-based model, in conjunction with a temporal attention layer, that exploits the available location information along with the obtained temperature-humidity data to generate the corresponding AQI label. Additionally, to mitigate the problem of the sparsity of devices, we also propose the design and development of low-cost AQMD that can serve as a potential alternative to the sophisticated AQMSs. To evaluate \ourmethod{}, we collect an in-house data using the developed low-cost hardware and further proof its generalizability using an additional publicly available dataset. The results obtained from this evaluation show that \ourmethod{} is capable of providing quality annotations and can improve over time as more and more training data becomes available from the deployed AQMSs (or AQMDs). Naturally, if properly pre-trained using available data from the existing infrastructures, \ourmethod{} can enable the residents of developing and under-developed countries to monitor their surrounding air quality using low-cost THMs.



\balance

\bibliographystyle{ACM-Reference-Format}
  \bibliography{ms}

\end{document}